\documentclass[a4paper,12pt,reqno]{amsart}
\usepackage[english]{babel}

\usepackage{a4}
\usepackage{graphicx}
\usepackage{amsthm}
\usepackage{amscd}
\usepackage{xpatch}

\usepackage{hyperref}
\usepackage{doi}

\usepackage{tikz}
\usepackage{tikz_graphs}
\usetikzlibrary{decorations.pathmorphing,shapes}
\usetikzlibrary{arrows.meta,calc}

\def\Xint#1{\mathchoice
	{\XXint\displaystyle\textstyle{#1}}%
	{\XXint\textstyle\scriptstyle{#1}}%
	{\XXint\scriptstyle\scriptscriptstyle{#1}}%
	{\XXint\scriptscriptstyle\scriptscriptstyle{#1}}%
	\!\int}
\def\XXint#1#2#3{{\setbox0=\hbox{$#1{#2#3}{\int}$}
		\vcenter{\hbox{$#2#3$}}\kern-.5\wd0}}

\def\dashint{\Xint-}

\makeatletter
\xpatchcmd{%
\@maketitle}{%
\ifx\@empty\authors \else \@setauthors \fi
}{%
  \ifx\@empty\authors \else \@setauthors \fi
  \ifx\@empty\addresses \else\@setaddresses\fi
 }{\typeout{Patch successful}}{\typeout{Patch failed}}
\def\enddoc@text{}
\makeatother

\newtheorem{definition}{Definition}
\newtheorem{example}[definition]{Example}
\newtheorem{theorem}[definition]{Theorem}
\newtheorem{lemma}[definition]{Lemma}

\newtheorem{corollary}[definition]{Corollary}
\newtheorem{proposition}[definition]{Proposition}

\makeatletter
\@addtoreset{definition}{section}
\@addtoreset{equation}{section}
\makeatother

\newcommand{\RNum}[1]
    {\MakeUppercase{\romannumeral #1}}

\newcommand{\G}{\mathcal{G}}
\DeclareMathOperator{\Res}{Res}

\def\mathScale{1}	
\def\miterLimit{1}	
\def\smallScale{0.7}	

\def\posZero{0}

\def\resZero{0.6}
\def\resOne{3.5}

\def\numGap{0.1}
\def\braceGap{1.8}
\def\posMiniGraph{1.4}

\def\BraceWidth{0.5}
\def\eqGraphs{2}

\def\repNKL{}
\def\repNK{}

\begin{document}

\title[From scalar fields on quantum spaces to BTR]{From scalar
  fields on quantum spaces to blobbed  topological recursion}

\author[J. Branahl]{Johannes Branahl\textsuperscript{1}}
\author[A. Hock]{Alexander Hock\textsuperscript{2}}
\author[H. Grosse]{Harald Grosse\textsuperscript{3}}
\author[R. Wulkenhaar]{Raimar Wulkenhaar\textsuperscript{1}}

\address{\textsuperscript{1}Mathematisches Institut der
  Westfälischen Wilhelms-Universit\"at \hfill \newline
Einsteinstr.\ 62, 48149 M\"unster, Germany \hfill \newline
{\itshape e-mail:} \normalfont
\texttt{j\_bran33@uni-muenster.de},
\texttt{raimar@math.uni-muenster.de}}

\address{\textsuperscript{2}Mathematical Institute, University of Oxford,
  \newline
  Andrew Wiles Building, Woodstock Road, OX2 6GG, Oxford, United Kingdom
  \newline
  {\itshape e-mail:} \normalfont \texttt{alexander.hock@maths.ox.ac.uk}}

\address{\textsuperscript{3}Fakult\"at für Physik, Universit\"at Wien
  \newline
  Boltzmanngasse 5, 1090 Wien, Austria \newline
  {\itshape e-mail:} \normalfont \texttt{harald.grosse@univie.ac.at}}

\begin{abstract}
  We review the construction of the $\lambda\phi^4$-model on
  noncommutative geometries via exact solutions of Dyson-Schwinger
  equations and explain how this construction relates via (blobbed)
  topological recursion to problems in algebraic and enumerative geometry.
\end{abstract}

\subjclass[2020]{81T75, 81Q80, 30F30, 05A15}

\keywords{Quantum field theory, Noncommutative geometry, Matrix models,
  (Blobbed) Topological recursion, Renormalisation, Ribbon graphs}

\maketitle

\section{Introduction}

Quantum field theories on noncommutative spaces appeared at the end
of the last century \cite{Grosse:1992bm,Doplicher:1994tu,Filk:1996dm,
  Grosse:1995ar}. These investigations and compactifications of M-theory on
the noncommutative torus
\cite{Connes:1997cr} motivated the perturbative renormalisation programme
of QFT on noncommutative geometries.
Whereas renormalisable at one-loop order \cite{Martin:1999aq,
  Krajewski:1999ja}, a new class of problems (UV/IR-mixing
\cite{Minwalla:1999px,Chepelev:1999tt}) was found at higher loop order.

Two of us (HG+RW) found a way to avoid the UV/IR-mixing problem
for scalar fields by understanding that it signals the generation
of another marginal coupling \cite{Grosse:2004yu,Grosse:2004wte}. 
This coupling corresponds to a harmonic
oscillator potential and implements a particular
duality under Fourier transform \cite{Langmann:2002cc}. 
The duality-covariant scalar model (with oscillator potential) is
perturbatively renormalisable \cite{Grosse:2003aj,Grosse:2004yu}.
Moreover, the $\beta$-function of the
coupling constant vanishes at the self-duality point
\cite{Grosse:2004by,Disertori:2006uy}. The proof \cite{Disertori:2006nq}
led to a new solution strategy first formulated by two of us 
(HG+RW) in \cite{Grosse:2009pa} and then extended in 
\cite{Grosse:2012uv}.
All these developments and results have been
reviewed previously in great detail
\cite{Szabo:2001kg,Wulkenhaar:2006si,Rivasseau:2007ab,SurveyNCG}.

This paper provides the first review of the enormous progress made during
the last three years. It started with the exact solution by one of us (RW)
with E.~Panzer \cite{Panzer:2018tvy} of the non-linear Dyson-Schwinger
equation found in \cite{Grosse:2009pa} for the case of 2-dimensional
Moyal space.  A renewed interest in higher planar correlation
functions \cite{DeJong} established a link to the Hermitian 2-matrix
model \cite{Eynard:2005iq} which has a non-mixed sector
that follows \emph{topological recursion}
\cite{Eynard:2016yaa}. This observation identified the key to
generalise \cite{Panzer:2018tvy} to a solution of all quartic matrix
models \cite{Grosse:2019jnv} by three of us (AH+HG+RW).  After initial
investigations of the algebraic-geometrical structure in
\cite{Schurmann:2019mzu}, three of us (JB+AH+RW) identified in
\cite{Branahl:2020yru} the
objects which obey (conjecturally; proved in the planar case
\cite{Hock:2021tbl}) \emph{blobbed topological recursion}
\cite{Borot:2015hna}, a systematic extension 
of topological recursion \cite{Eynard:2007kz}. Several properties of
these objects have already been investigated
\cite{Branahl:2020uxs}.

{\footnotesize\tableofcontents}

\section{What are quantum fields on quantum spaces?}

\label{sec:motiv}
\subsection{The free scalar field}

Quantum Physics was developed (by Planck, Heisenberg, Schr\"odinger
and others) between 1900 and 1926, special relativity by Einstein in
1905. Attempts to combine both led to the Klein-Gordon and the Dirac
equations. These equations, coupled to electromagnetic potentials,
describe energy levels of the electron and other particles.
Certain energy levels predicted to be degenerate are
split in nature (Lamb-shift). These tiny corrections are explained
by quantum electrodynamics (QED) and other quantum field theories.

A standard treatment in quantum field theory consists in expanding it
around a linear, exactly solvable, model. A favourite example is the
free scalar field in $D$ dimensions which arises by canonical
quantisation of the Klein-Gordon equation. The two-point function of the
free scalar field has an analytic continuation in time to a
two-point function on Euclidean space:
\begin{align}
G(t,\vec{x}) = \int \frac{d^{D-1}\vec p}{(2\pi)^{D-1} \cdot 2\omega_p}
e^{-\omega_p |t| +  \mathrm{i} \vec p \vec x },\qquad
\omega_p=\sqrt{M^2+\vec{p}^2}\;.
\label{Schwinger-2pt}
\end{align}
According to Minlos' theorem there exists a measure on the space $X$
of tempered distributions such that $G(t,\vec{x})= \int_X d \mu (\phi)
\; \phi(t,\vec{x})\phi(0,\vec{0})$, where $\phi(t,\vec{x})$ is a
stochastic field.
Moments of $d \mu (\phi)$ are understood as
Euclidean correlation functions.
They fulfil the
Osterwalder-Schrader (OS) axioms \cite{Osterwalder:1974tc}
of smoothness, Euclidean covariance,
OS positivity and symmetry.

\subsection{Interacting fields and renormalisation}

\label{sec:renorm}

The Minlos measure associated with (\ref{Schwinger-2pt})
is the starting point for attempts to construct
interacting models (in the Euclidean picture).
They are formally obtained by a deformation
\begin{align}
d\mu(\phi) \mapsto d\mu_{\mathrm{int}}(\phi)
= \frac{d\mu(\phi) \;e^{-S_{\mathrm{int}}(\phi)}}{
\int_X d\mu(\phi) \;e^{-S_{\mathrm{int}}(\phi)}},
\label{measure-int}
\end{align}
where the derivative 
of the functional $S_{\mathrm{int}}$
is non-linear in $\phi$.

As a matter of fact, in all cases of interest this deformation is
problematic because there exist (many) moments $\int_X d\mu_{\mathrm{int}}(\phi)
\;\phi(t_1,\vec{x}_1)\cdots \phi(t_n,\vec{x}_n)$ which diverge. To
produce meaningful quantities a procedure known as renormalisation
theory is necessary. Its first step is regularisation, which amounts
to understand the space $X$ of all $\phi$ as limit of a sequence (or
net) of finite-dimensional spaces $X_\alpha$. Then every
finite-dimensional space is endowed with its own functional
$S^\alpha_{\mathrm{int}}(\phi)$ which is carefully adjusted so that certain
moments $\int_{X_\alpha} d\mu_{\mathrm{int}}^\alpha(\phi) \;
\phi(x_1)\cdots \phi(x_n)$ stay constant when
increasing $\alpha$. Hence, they at least have a limit when
approaching $X$.  The challenge is to make sure that an adjustment of
finitely many moments suffices to render all moments meaningful; the
theory is then called renormalisable. In realistic
particle physics models, this was only achieved in
infinitesimal neighbourhoods of the free theory, which by far miss
the required physical parameter values. Much harder are models which
require adjustment of infinitely many moments to render infinitely many
other moments meaningful; Einstein gravity could be of this type.

\subsection{Scalar fields on quantum spaces}

This article reviews a framework of quantum field theory where the
renormalisation programme sketched in sec.~\ref{sec:renorm} can be
fully implemented for truly interacting fields. Our fields do not live
on familiar space-time; they live on a quantum space (a
\textit{quantised} space whose points follow non-trivial commutation
relations -- the main example in this work will be the noncommutative
Moyal space).  It is conceivable that such quantum spaces could be a
good description of our world when gravity and quantum physics are
simultaneously relevant \cite{Doplicher:1994tu}. It is probably
difficult to give a picture of a quantum space, but it is fairly easy
to describe scalar fields on it.

For that we let our finite-dimensional spaces $X_\alpha$ of
sec.~\ref{sec:renorm} be the spaces $H_N$ of Hermitean $N\times N$-matrices.
Given a sequence $(E_1,E_2,\dots)$ of positive real numbers, the energies, we
construct the Minlos measure $d\mu(\phi)$ of a free scalar field on
quantum space by the requirement
\begin{align}
\int_{H_N} d\mu(\phi) \;\phi_{kl}=0,\qquad
\int_{H_N} d\mu(\phi) \;\phi_{kl}\phi_{k'l'}
= \frac{\delta_{kl'}\delta_{lk'}}{N(E_k+E_l)},
\label{minlos-q}
\end{align}
and factorisation of higher $n$-point functions.
Here, the $(\phi_{kl})$ are the matrix elements of $\phi \in H_N$.
The $E_k$ should be viewed as eigenvalues of the Laplacian on our
quantum space. Their asymptotic
behaviour (for $N\to \infty$) defines a dimension of the quantum
space as the smallest
number $D$ such that $\sum_{k=1}^\infty E_k^{-D/2-\epsilon}$ converges
for all $\epsilon>0$.

We deform the Minlos measure (\ref{minlos-q}) as in
(\ref{measure-int}) via a quartic functional,
\begin{align}
d\mu_{\mathrm{int}}(\phi)=
\frac{d\mu(\phi) \;e^{-\frac{\lambda N}{4} \mathrm{Tr}(\phi^4)}}{
\int_{H_N} d\mu(\phi) \;e^{-\frac{\lambda N}{4} \mathrm{Tr}(\phi^4)}},\qquad
d\mu\text{ as in } (\ref{minlos-q}).
\label{minlos-q-int}
\end{align}
The deformation (\ref{minlos-q-int}) is a 
quartic analogue of the Kontsevich model \cite{Kontsevich:1992ti} in which
a cubic potential $\frac{\mathrm{i} N}{6} \mathrm{Tr}(\phi^3)$
is used to deform (\ref{minlos-q}). The Kontsevich model gives deep insight
into the moduli space $\overline{\mathcal{M}}_{g,n}$ of complex curves and
provides a rigorous formulation of quantum gravity in two dimensions
\cite{Witten:1990hr}. 
For obvious reasons we call the model which studies
(\ref{minlos-q-int}) and its moments the \emph{quartic Kontsevich model}.

In dimension $D\geq 2$ (encoded in the $E_k$), moments of
(\ref{minlos-q-int}) show the same divergences as discussed in
sec.~\ref{sec:renorm} on ordinary space-time. Between 2002 and 2004 we
treated them in a formal power series in (infinitesimal) $\lambda$. It
turned out that an affine rescaling $E_k \mapsto Z E_k + c$ was
enough, where $Z=Z(\lambda,N)$ and $c=c(\lambda,N)$ depend only on
$\lambda$ and the size $N$ of the matrices, but not on $k$. We
actually considered a more general Minlos measure where two further
renormalisation parameters $\Omega(\lambda,N)$ and
$\tilde{\lambda}(\lambda,N)$ were necessary \cite{Grosse:2004yu}; in
lowest $\lambda$-order we found \cite{Grosse:2004by} that
$\frac{\Omega^2}{\tilde{\lambda}}$ is independent of $N$. This was a
remarkable symmetry which indicated that the Landau ghost problem
\cite{Landau:1954??}  could be absent in this model.

This perspective influenced V.~Rivasseau who, with M.~Disertori,
R.~Gurau and J.~Magnen, proved in \cite{Disertori:2006nq} that the
model with quartic interaction functional (\ref{minlos-q-int})
tolerates the same $\lambda$ for all matrix sizes $N$, at least for
infinitesimal $\lambda$.  We understood immediately that their method
can potentially provide relations between moments of
$d\mu_{\mathrm{int}}(\phi)$.

This turned out to be true, with amazing consequences: The model was
revealed to be solvable. This solution has two aspects: First, the planar
2-point function of the measure (\ref{minlos-q-int}) satisfies a
closed non-linear equation \cite{Grosse:2009pa, Grosse:2012uv}. A
solution theory for this equation was developed in
\cite{Panzer:2018tvy, Grosse:2019jnv}; it suggested a particular
change of variables.  In a second step it was found in
\cite{Branahl:2020yru} that special combinations of the correlation
functions possess after complexification and the mentioned
change-of-variables a beautiful and universal algebraic-geometrical
structure: (blobbed) topological recursion.  The next section gives a
short introduction. We return to the model under consideration in
section~\ref{sec:solvingthemodel}.

\section{Algebraic geometrical structures}

\label{sec:AlgebraicGeometry}

\subsection{Topological recursion}
 
Topological recursion (TR) is a universal structure which is common to 
surprisingly many different topics in algebraic geometry,
enumerative geometry, noncommutative geometry, random matrix theory, string
theory, knot theory and more. It covers e.g.\
\textit{Witten's conjecture} \cite{Witten:1990hr} about intersection
numbers on the moduli space of stable complex curves
(proved by Kontsevich \cite{Kontsevich:1992ti}), Mirzakhani's
recursion \cite{Mirzakhani:2006fta}
for \textit{Weil-Petersson volumes} of bordered Riemann surfaces
and generating functions of \textit{Hurwitz numbers} \cite{Bouchard:2007hi}
with the same universal structure (see eq.\ (\ref{eq:tr}) below)!
The common structure was formulated by
B.~Eynard and N.~Orantin \cite{Eynard:2007kz} after insight into
the Hermitean 2-matrix model \cite{Chekhov:2006vd}. Since then it became
an active field of research. We refer to \cite{borot19} for an
overview covering more than 100 papers.

\medskip

Topological recursion constructs a family $\omega_{g,n}$ of symmetric
meromorphic differentials on products of Riemann surfaces $\Sigma$.
These $\omega_{g,n}$ are labeled by the genus $g$ and the number $n$
of marked points of a compact complex curve, they occur as invariants
of algebraic curves $P(x,y)=0$ (understood in parametric
representation $x(z)$ and $y(z)$).

\begin{figure}[h!t]
\centering
\includegraphics[width= 1.04\textwidth]{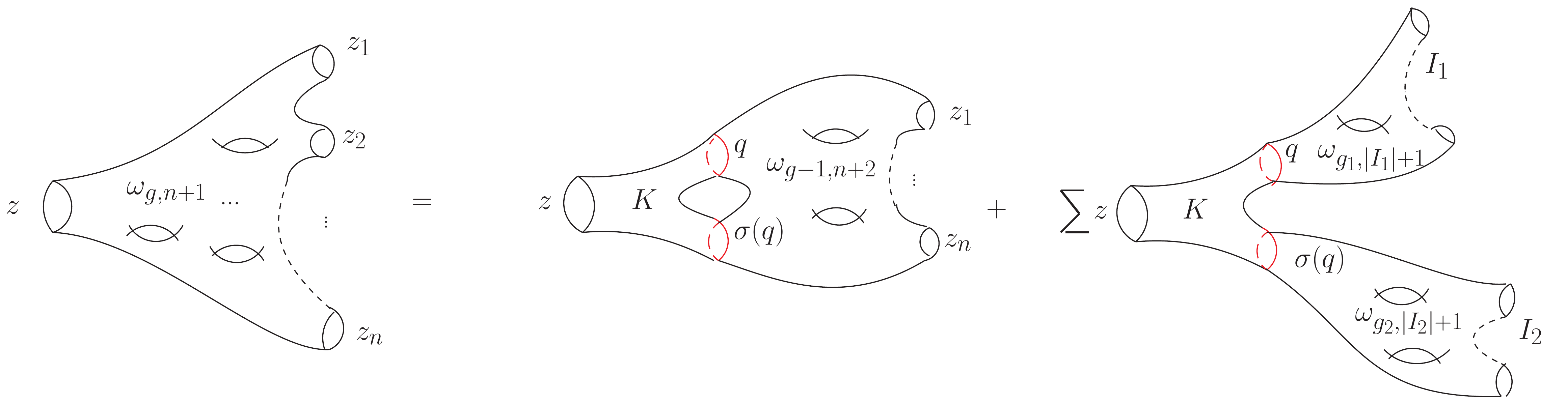} 
\caption{There are two different ways to obtain the left hand side
  $\omega_{g,n}$ by gluing the initial data (recursion kernel with
  three boundaries) with something of lower topology: Either one glues
  one object with one genus less and one boundary more $(g-1,n+2)$ with
  two boundaries of the kernel creating the missing genus, or one glues
  two objects with the kernel (causing no genus change): Then one has to
  account any $(g_1,n_1),(g_2,n_2)$ conform with the left hand side --
  the sum over all possible partitions in the master formula
  (\ref{eq:tr}) arises.}
	\label{fig:tr}
\end{figure}

From \textit{initial  data} consisting of a ramified covering
$x: \Sigma \to \Sigma_0$ of Riemann surfaces, 
a differential 1-form $\omega_{0,1}(z)=y(z)dx(z)$ and the \emph{Bergman kernel}
$\omega_{0,2}(z,u)=B(z,u)=\frac{dzdu}{(z-u)^2}$ 
(here assuming a genus-0 spectral curve),
TR constructs the meromorphic differentials $\omega_{g,n+1}(z_1,...,z_n,z)$ with
$2g+n\geq 2$ via the following universal formula
(in which we abbreviate $I=\{z_1,...,z_n\}$):
\begin{align}
\label{eq:tr}
& \omega_{g,n+1}(I,z)
  \\
  & =\sum_{\beta_i}
  \Res\displaylimits_{q\to \beta_i}
  K_i(z,q)\bigg(
  \omega_{g-1,n+2}(I, q,\sigma_i(q))
  +\hspace*{-1cm} \sum_{\substack{g_1+g_2=g\\ I_1\uplus I_2=I\\
            (g_1,I_1)\neq (0,\emptyset)\neq (g_2,I_2)}}
  \hspace*{-1.1cm} \omega_{g_1,|I_1|+1}(I_1,q)
  \omega_{g_2,|I_2|+1}(I_2,\sigma_i(q))\!\bigg)\;.
  \nonumber
\end{align}
This construction proceeds recursively 
in the negative Euler characteristic $-\chi=2g+n-2$. 
Here we need to define:
\begin{itemize}
\item A sum over the \textit{ramification points} $\beta_i$ of the ramified
  covering  $x:\Sigma\to \Sigma_0$, defined via $dx(\beta_i)=0$.
  \\\textbf{Example}: $x(z)=z^2$ for a Riemann surface $\Sigma=\hat{\mathbb{C}}
  :=\mathbb{C}\cup \{\infty\}$ with
  $\beta=0$. 
  The two sheets merge at $z=\beta=0$, but also at $z= \infty$ which is exceptional.
  
\item the \textit{local Galois involution} $\sigma_i\neq \mathrm{id}$
  defined via $x(q)=x(\sigma_i(q))$ near $\beta_i$ having the fixed
  point $\beta_i$ itself. \\
  \textbf{Example}: $x(z)=z^2$ gives $\sigma(z)=-z$;
\item the \textit{recursion kernel} $K_i(z,q)
  =\frac{\frac{1}{2}\int^{q'=q}_{q'=\sigma_i(q)}
    B(z,q')}{\omega_{0,1}(q)-\omega_{0,1}(\sigma_i(q))}$  constructed
  from the initial data. 
\end{itemize}
To orientate oneself within this jungle of definitions, we turn the
master formula into a picture (Fig.~\ref{fig:tr}). The recursion becomes a successive gluing of objects at their boundaries, starting
with the recursion kernel and two cylinders and then becoming
more and more complicated.

We conclude this subsection by giving the rather simple initial
data of the three previously listed prime examples ($\omega_{0,2}=B$
and $\Sigma = \hat{ \mathbb{C}}=\Sigma_0$ in all cases):
\begin{itemize}
\item \textbf{Witten's conjecture}: $x(z)=z^2$, $\omega_{0,1}(z)=2z^2dz$;
\item \textbf{Weil-Petersson volumes}: $x(z)=z^2$, $\omega_{0,1}(z)=\frac{4z}{\pi}\sin(\pi z)dz$;
\item \textbf{Simple Hurwitz numbers}: $x(z)=-z+\log (z)$, $\omega_{0,1}(z)=(1-z)dz$.
\end{itemize}

\subsection{Blobbed topological recursion}

We emphasised that topological recursion covers a large spectrum
of examples in enumerative geometry, 
mathematical physics, etc. The
model under consideration fits perfectly into an extension of TR 
developed in 2015 by G.~Borot and S.~Shadrin \cite{Borot:2015hna}:
\emph{Blobbed topological recursion (BTR)}. 

\smallskip

Its philosophy is quite analogous to that of TR, however the recursion
is equipped with an infinite stack of further initial data,
successively contributing to each recursion step. More precisely, the
meromorphic differentials
\begin{align*}
  \omega_{g,n}(...,z)=\mathcal{P}_z\omega_{g,n}(...,z)
  +\mathcal{H}_z\omega_{g,n}(...,z)
\end{align*}
decompose into a {polar part} $\mathcal{P}_z\omega_{g,n}$ (with poles
in a selected variable $z$ at ramification points) and a
\textit{holomorphic part} $\mathcal{H}_z\omega_{g,n}$ with poles
somewhere else. The polar part follows exactly the usual TR
\cite{Eynard:2007kz}, whereas the holomorphic part is not given via a
universal structure.
\begin{figure}[h!]
  \centering
    \includegraphics[width= 0.99\textwidth]{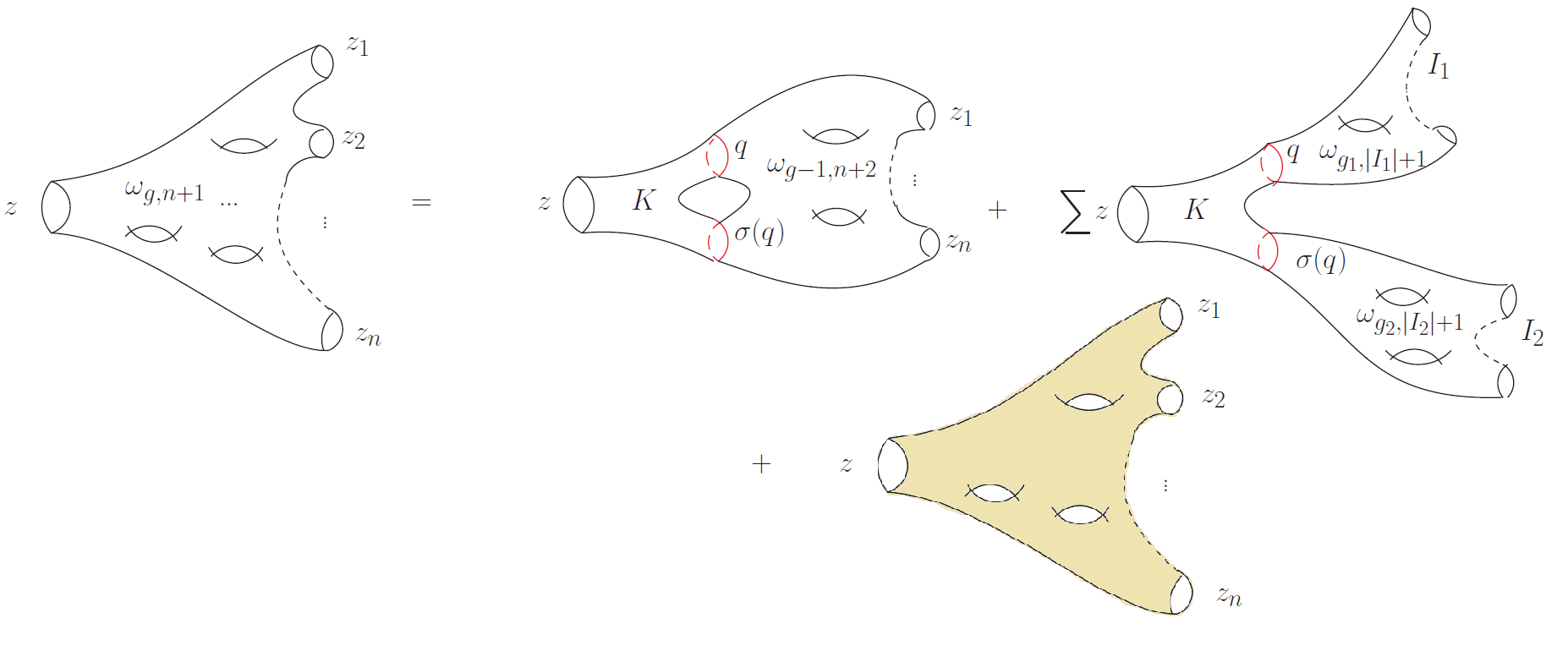} 
    \caption{Graphical interpretation of blobbed topological recursion: The usual recursion formula is enriched by a holomorphic (at ramification points) add-on $\mathcal{H}_z\omega_{g,n+1}$ (coloured) that appears as a surplus structure in the solution of the loop equations. It has to be seen as additional data that has to be taken into account at each further recursion step. 
      \label{diagrams}}
\end{figure}
This extended theory was baptised \textit{blobbed} due to the
occurring purely holomorphic parts (for $2g+m-2>0$)
$\phi_{g,m}(z_1,...,z_{m-1},z)= \mathcal{H}_{z_1}...
\mathcal{H}_{z_{m-1}}\mathcal{H}_z\omega_{g,m}(z_1...,z_{m-1},z)$,
called the \textit{blob}.  A family $\omega_{g,m}$ obeys BTR iff it
fulfils \textit{abstract loop equations} \cite{Borot:2013lpa}:
\begin{enumerate}
\item $\omega_{g,m}$ fulfils the so-called \textit{linear loop equation}
if
\begin{align}
\omega_{g,m+1}(u_1,...,u_m,z)+
\omega_{g,m+1}(u_1,...,u_m,\sigma_i(z))=
\mathcal{O}(z-\beta_i)dz
\label{lle}
\end{align}
is a holomorphic linear form for $z \to \beta_i$ with (at least) a simple zero
at $\beta_i$. 
\item $\omega_{g,m}$ fulfils the
\textit{quadratic loop equation} if
\begin{align}
Q^i_{g,m+1}&:=\omega_{g-1,m+2}(u_1,...,u_m,z,\sigma_i(z))
+ \hspace*{-0.5cm}
\sum_{\substack{g_1+g_2=g \\ I_1\uplus I_2=\{u_1,...,u_m\}}}
\hspace*{-0.8cm}
   \omega_{g_1,|I_1|+1}(I_1,z)
    \omega_{g_2,|I_2|+1}(I_2,\sigma_i(z))
\nonumber \\[-2ex]
&=\mathcal{O}((z-\beta_i)^2)(dz)^2
\label{qle}
\end{align}
is a holomorphic quadratic form with at least a double zero at $z \to \beta_i$. 
\end{enumerate}
Although these formulae seem simple, a proof that the actual
$\omega_{g,m}$ of a certain model fulfil the abstract loop equations
may demand some sophisticated techniques.
We list three models governed by BTR that were investigated within
the last years: 
\begin{itemize}
\item \textbf{Stuffed maps} \cite{Borot:2013fla}: The investigation of
  \textit{stuffed maps} arising from the multi-trace Hermitian
  one-matrix model -- not perfectly following the established theory of
  TR -- gave the motivation to formulate BTR in its full generality
  (two years later, \cite{Borot:2015hna}).

\item \textbf{Tensor models} \cite{Bonzom:2020xaf}: Tensor models are
  the natural generalisation of matrix models and are now known to be
  covered by BTR at least in the case of \textit{quartic melonic
    interactions} for arbitrary tensor models.

\item \textbf{Orlov–Scherbin partition functions}
  \cite{Bychkov:2020yzy}: Using $n$-point differentials corresponding to
  Kadomtsev-Petviashvili tau functions of hypergeometric type
  (Orlov–Scherbin partition functions) that follow BTR, the authors
  were able to reprove previous results
  and also to establish new enumerative
  problems in the realm of Hurwitz numbers.
\end{itemize}
A fourth example is the quartic interacting quantum field theory
defined by the measure (\ref{minlos-q-int}).

\section{Solving the Model}
\label{sec:solvingthemodel}

\subsection{The setup}

Moments of the measure $d\mu_{\mathrm{int}}(\phi)$ defined in (\ref{minlos-q-int})
come with an intricate substructure. They first decompose into
connected functions (or cumulants)
\begin{align}
\int_{H_N} d\mu_{\mathrm{int}}(\phi)\;\phi_{k_1l_1}\cdots \phi_{k_nl_n}
=\sum_{\substack{\text{partitions }\\
\mathcal{P} \text{ of } \{1,\dots,n\} }}
\prod_{\text{blocks }\mathcal{B} \in \mathcal{P}} \Big\langle
\prod_{i\in \mathcal{B}} \phi_{k_il_i} \Big\rangle_c.
\label{partition}
\end{align}
Because of the invariance of (\ref{minlos-q-int}) under $\phi\mapsto
-\phi$, the only contributions come from $n$ and all
$|\mathcal{B}|$ even.
Take all $k_1,...,k_n$ pairwise different. Then it follows from
(\ref{minlos-q}) that contributions to cumulants
$\big\langle
\phi_{k_1l_1} \cdots \phi_{k_nl_n} \big\rangle_c$
vanish unless the $l_i$ are a permutation of the $k_j$. Any permutation
is a product of cycles, and after renaming matrix indices, only
cumulants of the form
\begin{align}
&N^{2-b} G_{|k_1^1\dots k_{n_1}^1|\dots |k_1^b\dots k_{n_b}^b|}
:=N^n
\Big\langle
\prod_{j=1}^b\prod_{i=1}^{n_j}
\phi_{k_i^j k_{i+1}^j} \Big\rangle_c
\label{eq:cumulants}
\end{align}
arise, where $k_{n_j+1}^j=k_1^j$ is cyclic.
Finally, the $G_{\dots}$ come with a grading, which is called
genus because it relates to the genus of Riemann surfaces:
\begin{align}
G_{|k_1^1\dots k_{n_1}^1|\dots |k_1^b\dots k_{n_b}^b|}
=\sum_{g=0}^\infty N^{-2g}
G^{(g)}_{|k_1^1\dots k_{n_1}^1|\dots |k_1^b\dots k_{n_b}^b|}.
\label{genus-exp}
\end{align}

These $G^{(g)}_{\dots}$ are not independent. They are related by
quantum equations of motions, called Dyson-Schwinger equations. Moreover,
symmetries of the model give rise to a Ward-Takahashi identity
\cite{Disertori:2006nq}
\begin{align}
0=\sum_{p=1}^{N}
\Big((E_k-E_l) \frac{\partial^2 }{\partial
  J_{kp} \partial J_{pl}}
-J_{lp} \frac{\partial}{\partial J_{kp}}+
J_{pk} \frac{\partial}{\partial J_{pl}}\Big)
\int_{H_N} d\mu_{\mathrm{int}}(\phi) \;e^{\mathrm{i} \mathrm{Tr}(\phi J)}
.
\label{WI-sum}
\end{align}
It breaks down to further relations between the $G^{(g)}_{\dots}$.
All these relations together imply a remarkable pattern
\cite{Grosse:2012uv}: The functions $G^{(g)}_{\dots}$ come with a
partial order, i.e.\ either two (different) functions are independent,
or precisely one is strictly smaller than the other.  The relations
respect this partial order: A function of interest depends only on
finitely many smaller functions. The smallest function is the planar
two-point function $G^{(0)}_{|ab|}$ which satisfies a closed
non-linear equation \cite{Grosse:2009pa}. The non-linearity makes this
equation hard to solve. The solution eventually succeeded with
techniques from complex geometry. First, the equation
extends to an equation for a holomorphic function
$G^{(0)}(\zeta,\eta)$.  Let $(e_1,...,e_d)$ be the pairwise different
values in $E_1,...,E_N$, which arise with multiplicities
$(r_1,...,r_d)$. To deal with the renormalisation problem 
in the limit $N,d\to \infty$, we have to rescale and shift these
values to $e_k \mapsto Z(e_k+\frac{\mu_{bare}^2}{2})$. 
Then $G^{(0)}_{|ab|}=G^{(0)}(e_a,e_b)$ where
$G^{(0)}(\zeta,\eta)$ satisfies the non-linear closed equation
\begin{align}
&(\zeta+\eta+\mu^2_{bare}) Z G^{(0)}(\zeta,\eta)
\label{eq:Gcomplex}
\\
&=1-\frac{\lambda}{N}
\sum_{k=1}^{d}r_k
\Big( ZG^{(0)}(\zeta,\eta) \;ZG^{(0)}(\zeta,e_k)
- \frac{ZG^{(0)}(e_k,\eta) -ZG^{(0)}(\zeta,\eta) }{e_{k}-\zeta}\Big) .
\nonumber
\end{align}
For finite $d$ one can safely set $\mu_{bare}^2=0=Z-1$. We already
included $\mu_{bare}^2,Z$ in (\ref{eq:Gcomplex}) to prepare 
 the limit $N,d\to \infty$ in which, depending on the growth rate
of $r_k$, this equation will suffer from a divergence problem. One
has to carefully adjust
$Z(N,\lambda)$ and $\mu^2_{bare}(N,\lambda)$ to make 
$\lim_{N \to \infty}G^{(0)}(\zeta,\eta)$ well-defined.

\begin{figure}[h!]
  \centering
    \includegraphics[width= 0.99\textwidth]{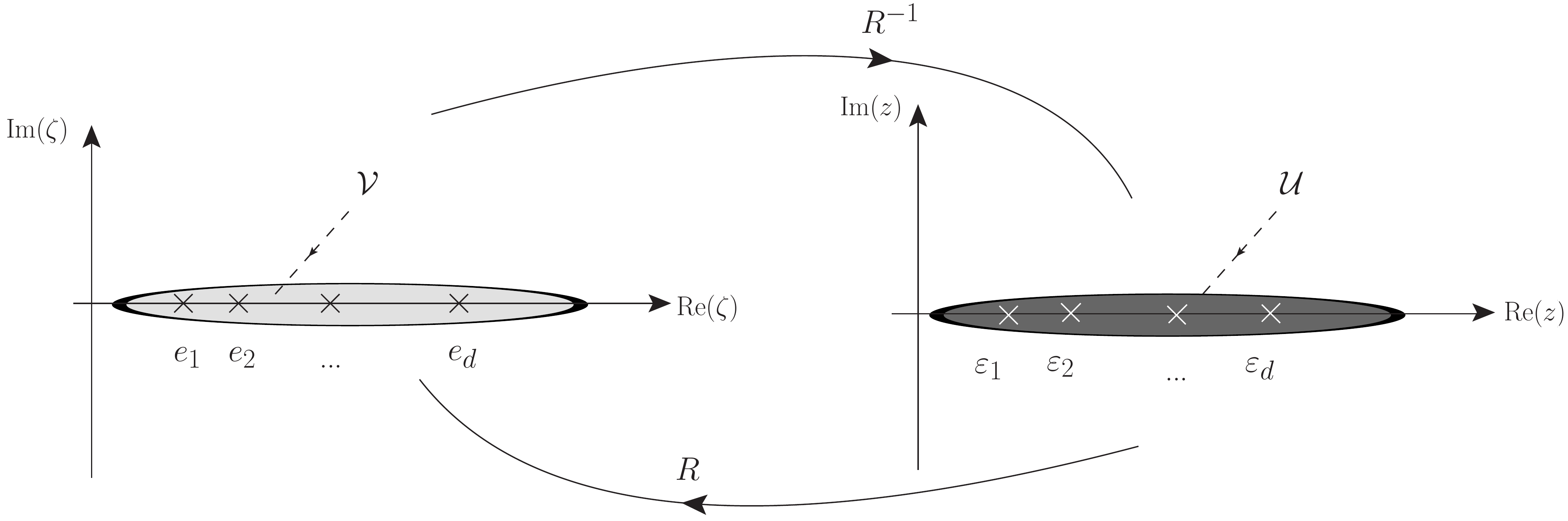} 
    \caption{Illustration of the change of variables: The
      biholomorphic map $R:\mathcal{U}\to \mathcal{V}$ with
      $R(\varepsilon_k)=e_k$ will later be enlarged
      to a ramified cover $R:\hat{\mathbb{C}}\to \hat{\mathbb{C}}$.
      Functions on $\mathcal{U}$ will meromorphically continue
      to the Riemann sphere $\hat{\mathbb{C}}=\mathbb{C}\cup\{\infty\}$.
      \label{fig:complexification}}
  \end{figure}

The key step to solve (\ref{eq:Gcomplex}) is a transform of variables
$\zeta\mapsto z=R^{-1}(\zeta)$ implemented by a biholomorphic
mapping $R^{-1}$ depicted in Figure~\ref{fig:complexification}.
For appropriately chosen
preimages $R(\varepsilon_k)=e_k$ we introduce 
another holomorphic function
$\mathcal{G}^{(0)}$ by $\mathcal{G}^{(0)}(z,w)=G^{(0)}(R(z),R(w))$.
We require that $R$ and $\mathcal{G}^{(0)}$ relate by
\begin{align}
R(z)+\mu^2_{bare}+\frac{\lambda}{N}
\sum_{k=1}^{d} r_k Z\mathcal{G}^{(0)}(z,\varepsilon_k)
+\frac{\lambda}{N}\sum_{k=1}^{d}
\frac{r_k}{R(\varepsilon_k)-R(z)}=-R(-z).
\label{fG-ansatz}
\end{align}
These steps turn (\ref{eq:Gcomplex}) into
\begin{align}
&(R(w)-R(-z))
Z\mathcal{G}^{(0)}(z,w)
=1+\frac{\lambda}{N}
\sum_{k=1}^{d}r_k
\frac{Z\mathcal{G}^{(0)}(\varepsilon_k,w)}{R(\varepsilon_k)-R(z)} ,
\label{eq:fG}
\end{align}
which is a \emph{linear} equation for which a solution theory exists.
It expresses $\mathcal{G}^{(0)}$ in terms of the not yet known
function $R$.
Inserting it into (\ref{fG-ansatz}) yields a complicated equation for $R$.
The miracle is that relatively  mild assumptions on $R$
allow to solve this problem:
\begin{theorem}[\cite{Schurmann:2019mzu}, building on \cite{Grosse:2019jnv}] 
\label{throm1}
Let $\lambda,e_k>0$ and $\mu^2_{bare}=0=Z-1$
(absent renormalisation). 
Assume that there is a rational function 
$R:\hat{\mathbb{C}}\to \hat{\mathbb{C}}$ with
\begin{enumerate}
\item $R$ has degree $d+1$, is normalised to 
  $R(\infty)=\infty$ and biholomorphically maps a domain $\mathcal{U}
  \subset \mathbb{C}$ to a neighbourhood 
$\mathcal{V}$ of a real interval that contains $e_1,\dots,e_d$.

\item $R$ satisfies \eqref{fG-ansatz} with $\mathcal{G}^{(0)}$
  the solution of
\eqref{eq:fG}, where $z,w,\varepsilon_k\in \mathcal{U}$.
\end{enumerate}
Then
the functions 
$R$ and $\mathcal{G}^{(0)}$ are uniquely identified as
\begin{align}
R(z)&=z-\frac{\lambda}{N} \sum_{k=1}^d \frac{\varrho_k}{\varepsilon_k+z}\;,\qquad
R(\varepsilon_k)=e_k\;,\quad
\varrho_k R'(\varepsilon_k)=r_k\;,
\label{R}
\\
\mathcal{G}^{(0)}(z,w)&=\frac{\displaystyle 
1 -\frac{\lambda}{N} \sum_{k=1}^d \frac{r_k}{
(R(z)-R(\varepsilon_k))(R(\varepsilon_k)-R({-}w))}
\prod_{j=1}^d \frac{
R(w){-}R({-}\widehat{\varepsilon_k}^j)}{ R(w)-R(\varepsilon_j)}
}{R(w)-R(-z)}\;.
\label{Gzw-final}
\end{align}
Here, the solutions of $R(v)=R(z)$ 
are denoted by $v\in\{z,\hat{z}^1,\dots,\hat{z}^d\}$ with $z\in \mathcal{U}$
when considering $R:\hat{\mathbb{C}}\to \hat{\mathbb{C}}$. The symmetry
$\mathcal{G}^{(0)}(z,w)=\mathcal{G}^{(0)}(w,z)$ is automatic.
\end{theorem}
\noindent We discuss later the renormalisation problem
$\mu_{bare}^2(N,\lambda)\neq 0 \neq Z(N,\lambda)-1$ in the limit
$N\to \infty$.

The change of variables (\ref{R}) identified in the solution
(\ref{Gzw-final}) of the two-point function
is the starting point for everything else. However,
although the equations for all other $G_{..}$ are affine, they
cannot be solved directly. It was understood in \cite{Branahl:2020yru}
that one
has first to introduce two further families of functions,
as we will explain in the next subsection.

\subsection{BTR of the Quartic Kontsevich Model}\label{Sec.BTR}

\begin{figure}[h!t]
  \centering
    \includegraphics[width= 0.99\textwidth]{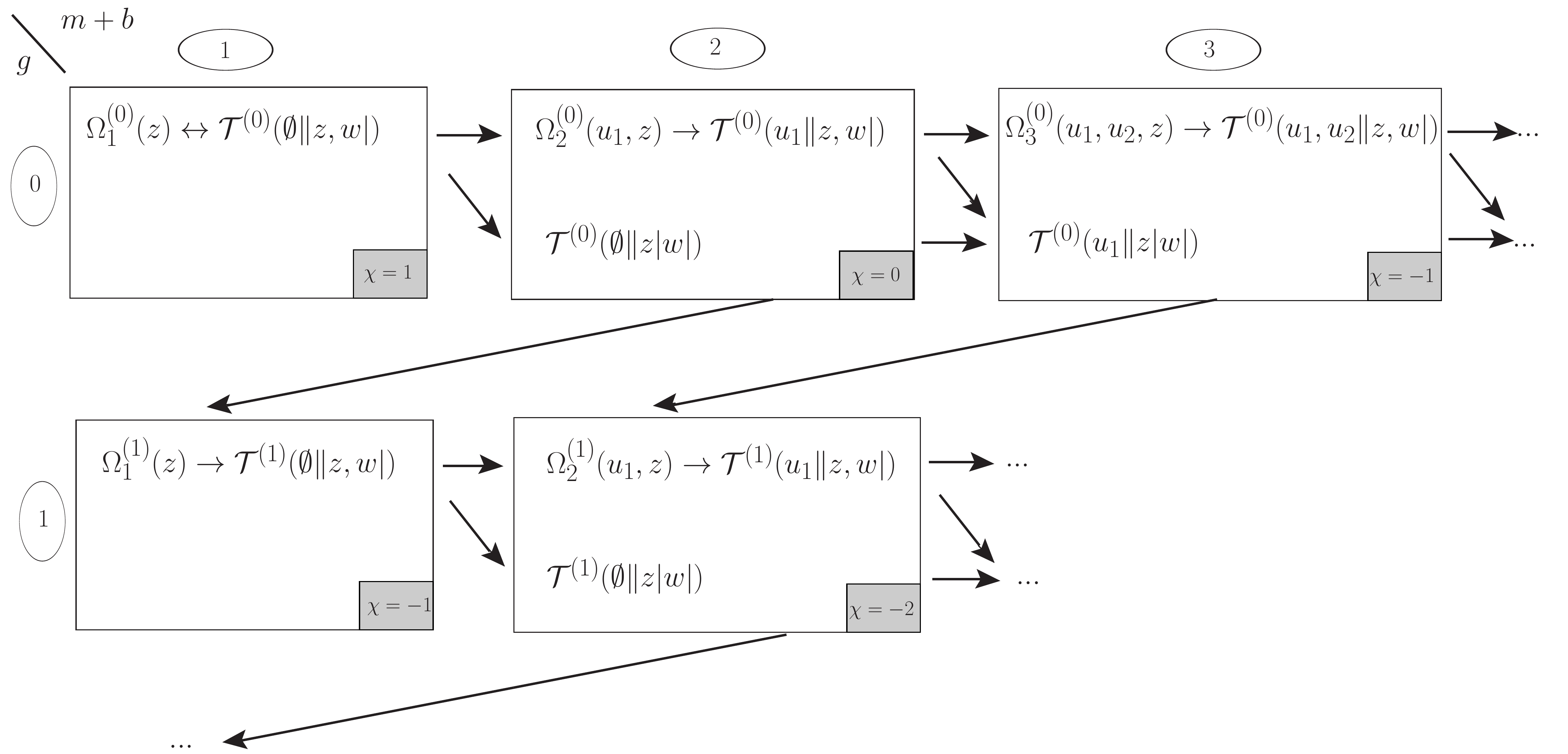} 
    \caption{Illustration of the interwoven solution procedure, ordered by $-\chi$.
    Theorem \ref{throm1} which simultaneously gives
    $ \Omega^{(0)}_1(z)=\frac{1}{N}\sum_{k}r_k\G^{(0)}(\varepsilon_k,z)$ and the
    two-point function $\mathcal{T}^{(0)}(\emptyset\|z,w) =
    \mathcal{G}^{(0)}(z,w)$ is the starting point.
    A generic box at position $(g,m+b)$ contains $\Omega_m^{(g)}(u_1,...,u_{m-1},z)
    \to \mathcal{T}^{(g)}(u_1,...,u_{m-1}\|z,w|)$ and $\mathcal{T}^{(g)}(u_1,...,u_{m-2}\|z|w|)$.
    \label{fig:euler}}
\end{figure}

To give an impression how blobbed topological recursion
is related to our model, we first introduce the aforementioned
three families of functions whose importance became clear
during an attempt of solving the $2{+}2$-point function:
\begin{itemize}
\item the cumulants 
  $G^{(g)}_{|k_1^1\dots k_{n_1}^1|\dots |k_1^b\dots k_{n_b}^b|}$
  defined in (\ref{eq:cumulants}), (\ref{partition}) and
  (\ref{genus-exp}), called
  $(n_1{+}...{+}n_b)$-point functions of genus $g$;
  
\item  the \textit{generalised correlation functions} defined as derivatives 
(here we need $E_k=e_k$ and $r_k=1$; this restriction is later relaxed)
  \begin{align}
T^{(g)}_{q_1,q_2,...,q_m\|k_1^1...k_{n_1}^1|k_1^2...k_{n_2}^2|...|k_1^b...k_{n_b}^b|}
:=\frac{(-N)^m\partial^m}{\partial e_{q_1}\partial e_{q_2}...\partial 
e_{q_m}}G^{(g)}_{|k_1^1...k_{n_1}^1|k_1^2...k_{n_2}^2|...|k_1^b...k_{n_b}^b|}\;;
\end{align}
\item functions $\Omega_{q_1,...,q_m}$ recursively defined by 
\begin{align}
\Omega^{(g)}_{q_1,...,q_m} &:= 
\frac{(-N)^{m-1}\partial^{m-1}\Omega^{(g)}_{q_1}}{
\partial e_{q_2}...\partial 
e_{q_{m}}} +\frac{\delta_{m,2}\delta_{g,0}}{(e_{q_1}-e_{q_2})^2}\qquad
\text{for } m\geq 2
\label{eq:Omega-gm}
\end{align}
and
$\Omega^{(g)}_{q} := \frac{1}{N}\sum_{k=1}^N G^{(g)}_{|qk|}
+\frac{1}{N^2}G^{(g)}_{|q|q|}$. They are symmetric in their indices
and will soon be related to \textit{meromorphic differentials which
  satisfy BTR}.

\end{itemize}
A straightforward extension gave birth to
\textit{(generalised) Dyson-Schwinger equations} between these functions.
They first extend to equations
for holomorphic functions on several copies of $\mathcal{V}$ and then via
the change of variables $R,R^{-1}$ to equations between meromorphic functions
\begin{align}
&\mathcal{G}^{(g)}(z_1^1,\dots, z_{n_1}^1|\dots |z_1^b,\dots z_{n_b}^b)\;,\quad
\mathcal{T}^{(g)}(u_1,...,u_m\|
z_1^1,\dots, z_{n_1}^1|\dots |z_1^b,\dots z_{n_b}^b|)\;,
\nonumber
\\
&\Omega^{(g)}_m(u_1,...,u_m)
\label{families}
\end{align}
on several copies of $\hat{\mathbb{C}}$. It was shown in
\cite{Branahl:2020yru} that these equations can be
recursively solved in a triangular pattern
of interwoven loop equations connecting
these three families $(\mathcal{G}^{(g)},\mathcal{T}^{(g)},\Omega^{(g)}_m)$ of
(\ref{families}), see Figure~\ref{fig:euler}.
The arrows represent very different difficulties. It is easy to express
every next $\mathcal{T}^{(g)}$ in terms of previous $\Omega^{(g')}_{m}$,
but the result is an extremely lengthy and complicated equation for the
next $\Omega^{(g)}_{m}$ in terms of the previous $\Omega^{(g')}_{m'}$.
Obtaining a $(n_1{+}...{+}n_{b})$-point function of genus $g$ from
$\mathcal{T}^{(g')},\Omega^{(g')}_{m'}$ in all boxes with $g'\leq g$ and
$m'+b'\leq b$ is also easy (unless one wants to make the symmetries manifest).

To our enormous surprise, the solution of the first of these very difficult
equations for $\Omega^{(g)}_m$ with $2g+m-2\geq 0$
turned out to be ravishingly simple and structured.
After the solution of $\Omega_m^{(g)}$ for
$(g,m)\in\{(0,2),(0,3),(0,4),(1,1)\}$ in \cite{Branahl:2020yru}
via the interwoven equations, it
became nearly obvious that the meromorphic differentials
$\omega_{g,n}$ defined by
\begin{align}
\omega_{g,n}(z_1,...,z_n):=\lambda^{2-2g-n}\Omega^{(g)}_{n}(z_1,...,z_n)
dR(z_1)\cdots dR(z_n)
\end{align}
obey BTR. In this process
the variable transform $R$ is again of central importance. It
provides the spectral curve 
$(x:\hat{\mathbb{C}}\to \hat{\mathbb{C}},
\omega_{0,1}=ydx,\omega_{0,2})$ discussed in 
Sec.~\ref{sec:AlgebraicGeometry} with
\begin{align}\label{specurve}
  x(z)=R(z)\;,\quad
  y(z)=-R(-z)\;,\quad
  \omega_{0,2}(u,z)=\frac{du\,dz}{(u-z)^2}+\frac{du\,dz}{(u+z)^2}\;.
\end{align}
We underline the appearance of some additional initial data
in $\omega_{0,2}$, namely $\frac{du\,dz}{(u+z)^2}=-B(u,-z)$
(Bergman kernel with one changed sign).

The next steps will consist in identifying structures and equations
directly for the family $\omega_{g,n}$, avoiding the
$\mathcal{T}^{(g)}$. This task was accomplished for the planar sector 
$\omega_{0,n}$ in \cite{Hock:2021tbl}. The symmetry of the spectral
curve, $y(z)=-x(-z)$ and $\omega_{0,2}(u,z)=B(u,z)-B(u,-z)$
played a key r\^ole. This symmetry extends to a deep involution identity
\begin{align}
&  \omega_{0,|I|+1}(I,q)
  +\omega_{0,|I|+1}(I,- q)
\label{eq:flip-om}
\\
&=-\sum_{s=2}^{|I|} \sum_{I_1\uplus ...\uplus I_s=I}
\frac{1}{s} \Res\displaylimits_{z\to q}  \Big(
\frac{dR(-q) dR(z)}{(R(-z)-R(-q))^{s}}  \prod_{j=1}^s
\frac{\omega_{0,|I_j|+1}(I_j,z)}{dR(z)} 
\Big)\;.
\nonumber
\end{align}
With considerable combinatorial effort it was possible to prove
that this involution identity completely determines the moromorphic
differentials $\omega_{0,n+1}$ 
to the following structure astonishingly similar to usual TR:
\begin{theorem}[\cite{Hock:2021tbl}]\label{thmBTRplan} 
  Assume that $z\mapsto \omega_{0,n+1}(u_m,...,u_m,z)$ is for $m\geq 2$
  holomorphic at
  $z=-\beta_i$ and $z=u_k$ and has poles at most in points where the
  rhs of \eqref{eq:flip-om} has poles. Then equation \eqref{eq:flip-om} is
  for $I=\{u_1,...,u_m\}$ with $m\geq 2$ uniquely solved by
\begin{align}
  \omega_{0,|I|+1}(I,z)
  &= \sum_{i=1}^r
\Res\displaylimits_{q\to \beta_i}K_i(z,q)
  \sum_{I_1\uplus I_2=I} \omega_{0,|I_1|+1}(I_1,q)\omega_{0,|I_2|+1}(I_2,\sigma_
i(q))
  \label{sol:omega}
  \\
  &-\sum_{k=1}^m d_{u_k}
 \Big[\Res\displaylimits_{q\to - u_k}
\sum_{I_1\uplus I_2=I}
\tilde{K}(z,q,u_k)
d_{u_k}^{-1}\big( \omega_{0,|I_1|+1}(I_1,q)
\omega_{0,|I_2|+1}(I_2,q)\big)\Big]\,.
\nonumber
\end{align}
Here $\beta_1,...,\beta_r$ are the ramification points of the
ramified covering $R:\hat{\mathbb{C}}\to \hat{\mathbb{C}}$
given in \eqref{R} and
$\sigma_i\neq \mathrm{id}$ denotes the local Galois involution
in the vicinity of $\beta_i$, i.e.\ $R(\sigma_i(z))=R(z)$,
$\lim_{z\to \beta_i}\sigma_i(z)=\beta_i$.
By $d_{u_k}$ we denote the exterior differential in $u_k$, which on 1-forms
has a right inverse
$d^{-1}_u \omega(u)=\int_{u'=\infty}^{u'=u}\omega(u')$.
The recursion kernels are given by
\begin{align}
K_i(z,q)&:=   \frac{\frac{1}{2} (\frac{dz}{z-q}-\frac{dz}{z-\sigma_i(q)})
}{dR(\sigma_i(q))(R(-\sigma_i(q))-R(-q))}\;,\qquad
\nonumber
\\
\tilde{K}(z,q,u)&:=
\frac{\frac{1}{2}\big(\frac{dz}{z-q}-\frac{dz}{z+u}\big)}{dR(q)
  (R(u)-R(-q))}\;.
\label{eq:kernel}
\end{align}
The solution \eqref{sol:omega}$+$\eqref{eq:kernel}
coincides with the solution of the interwoven loop equations
depicted in Fig.~\ref{fig:euler}.
The linear and quadratic loop equations \eqref{lle} and \eqref{qle} hold.
The symmetry 
of the rhs of \eqref{eq:flip-om} under $q\mapsto -q$ is automatic.
\end{theorem}

To give an explicit formula for the holomorphic part as well, using
the structure of topological recursion itself, is to the best of our
knowledge exceptional. Higher genera $g$ are under current
investigation and require again an arduous solution of the interwoven
loop equations by hand. These non-planar symmetric meromorphic
differentials have an additional pole of higher order at the fixed
point $q=0$ of the involution $q \to -q$.

\begin{figure}[h!]
  \centering
  \includegraphics[width= 0.99\textwidth]{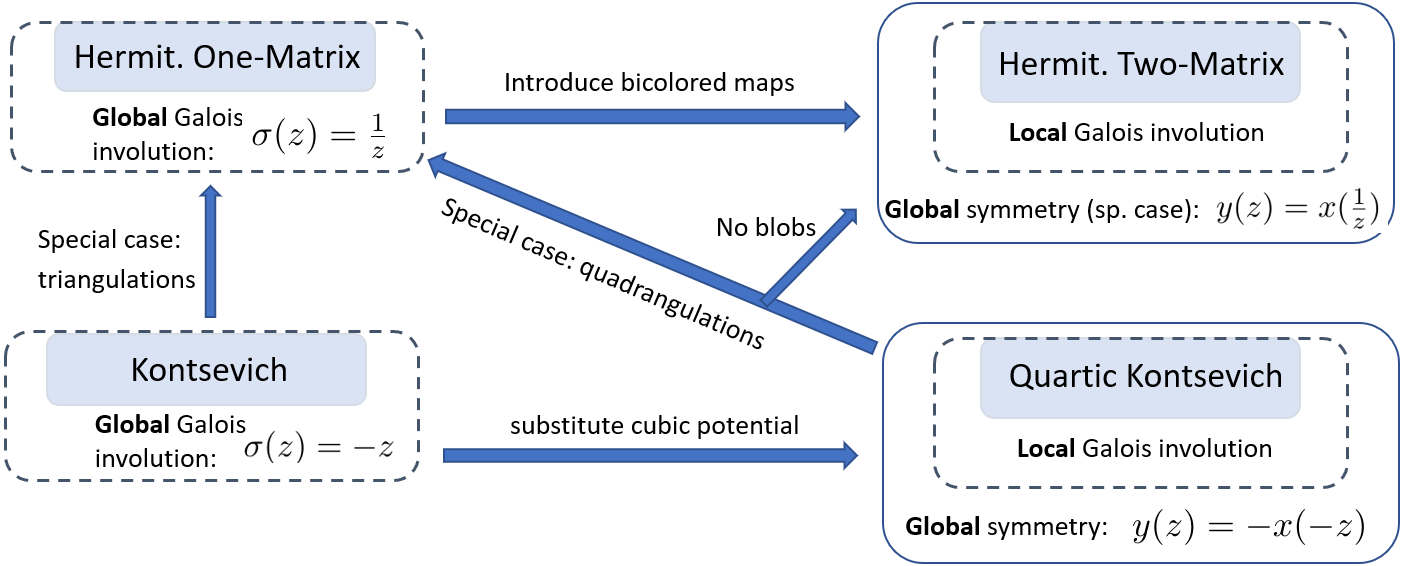} 
    \caption{Despite completely different mathematical structures behind the models, we can reach the limit both of the Hermitian one- and two-matrix model. Moreover, we stress that the global Galois involution of the one-matrix model and the Kontsevich model turn into the global symmetry of the spectral curve belonging to their more intricate siblings. \label{othermm}}
 \end{figure}

Figure~\ref{othermm} represents similarities between the quartic
Kontsevich model, the original Kontsevich model
\cite{Kontsevich:1992ti}, the Hermitian one- and two matrix models
\cite{Eynard:2016yaa,Eynard:2002kg,Chekhov:2006vd}).
 Although the modifications to go from one model
 to the other seem mild, the mathematics of the four
 models differs drastically. But they all fit into some flavour of
 topological recursion so that there is a fruitful exchange of methods.

\subsection{Solution strategy of all quartic models}

In the two previous subsections we assumed finite matrices,
in particular a truncated energy spectrum at finite $E_N$.
The relations to ordinary QFTs appear in a limit $N\to \infty$
and depend on the \textit{(spectral) dimension} encoded in the $E_k$.
In this subsection we describe this limit process. Note that the limit
$N\to \infty$ turns rational functions into 
transcendental functions so that most algebraic structures get lost.
Future research projects will address the questions 
whether parts of (B)TR survive and whether these surviving algebraic
structures are compatible with renormalisation. This subsection addresses
the simplest topological sector $(g,n)=(0,1)$.

Introducing the measure
$\varrho_0(t):=\frac{1}{N}\sum_{k=1}^dr_k\delta(t-e_k)$ we can turn the 
non-linear equation 
\eqref{eq:Gcomplex} into the integral equation
\begin{align}
&\bigg[\zeta+\eta+\mu^2_{bare}+\lambda\dashint_0^{\Lambda^2}dt\,\varrho_0(t)\bigg(ZG^{(0)}(\zeta,t)+\frac{1}{t-\zeta}\bigg)\bigg]Z G^{(0)}(\zeta,\eta)
\label{eq:GcomplexCont}
\\
&=1+\lambda\dashint_0^{\Lambda^2}
dt\,\varrho_0(t)
\Big(
 \frac{ZG^{(0)}(t,\eta)  }{t-\zeta}\Big) ,
\nonumber
\end{align}
where $\dashint$ is the Cauchy principle value
and $\Lambda^2=\mathrm{max}_{k}(e_k)$.
The limit $N\to \infty$ will be achieved in two steps.
In the first step we interpret 
the measure $\varrho_0$ as a H\"older-continuous function.
The renormalisation constants $Z,\mu^2_{bare}$ obtain a dependence
on $\Lambda^2$ which in the second step is sent to $\infty$.
Furthermore, the spectral dimension $D$ is also
provided by an integral representation depending on the
asymptotics of $\varrho_0(t)$ for $\Lambda^2\to \infty $,
being the smallest $D$ such that the integral
\begin{align}\label{specdim}
  \sum_{k=1}^{\infty} E_k^{-D/2-\epsilon}
  =\lim_{\Lambda^2\to\infty}\int_0^{\Lambda^2} dt\,
  \frac{\varrho_0(t)}{(1+t)^{D/2+\epsilon}}.
\end{align}
converges for all $\epsilon>0$.
\begin{example} 
\begin{itemize}
\item For an asymptotically constant measure $\varrho_0(t)\sim const$,
  the spectral dimension becomes $D=2$. The integral on the lhs of
  \eqref{eq:GcomplexCont} diverges logarithmically in $\Lambda^2$,
  which can be absorbed by $\mu^2_{bare}(\Lambda^2)$. The field
  renormalisation is set to $Z=1$.
\item For an asymptotically linear measure $\varrho_0(t)\sim t$, the
  spectral dimension becomes $D=4$. The renormalisation constants
  $\mu^2_{bare}(\Lambda^2)$ and $Z(\Lambda^2)$ have to be adapted
  carefully such that $\lim_{\Lambda^2 \to\infty}G^{(0)}(\zeta,\eta)$
  converges.
\item For a measure with asymptotic behaviour $\varrho_0(t)\sim t^a$
  and $a\geq 2$, the spectral dimension becomes $D\geq 6$. In this
  case, the model is not renormalisable anymore.
\end{itemize}
\end{example}

Assuming that the expression in the square brackets in
\eqref{eq:GcomplexCont} is known, then the
powerful solution theory of singular integral equations
\cite{Tricomi85} provides the explicit expression \cite{Grosse:2012uv}
\begin{align*}
  ZG^{(0)}(a,b)=\frac{e^{\mathcal{H}^\Lambda_a[\tau_b(\bullet)]\sin\tau_b(a)}}{
    \lambda \pi \varrho_0(a)}\qquad a,b\in [0,\Lambda^2],
\end{align*}
where
$\mathcal{H}^\Lambda_a[f(\bullet)]:=\frac{1}{\pi}\dashint_{0}^{\Lambda^2}\frac{dt
  \,f(t)}{t-a}$ is the finite Hilbert transform. Inserting this ansatz
into (\ref{eq:GcomplexCont}) gives a consistency equation
\cite{Panzer:2018tvy,Grosse:2019jnv} for the
angle function
$\tau_a:(0,\Lambda^2)\to [0,\pi]$ for $\lambda>0$ and
$\tau_a:(0,\Lambda^2)\to [-\pi,0]$ for $\lambda<0$:
\begin{align}\label{selfcont}
	\frac{\cot \tau_a(p)}{\lambda \pi \varrho_0(p)}=a+p+\mu^2_{bare}(\Lambda^2) +\lambda\pi \mathcal{H}_p^\Lambda [\varrho_0(\bullet)]+\frac{1}{\pi}\int_0^{\Lambda^2}dt\, \tau_p(t).
\end{align}
The solution of the angle function was first found in
\cite{Panzer:2018tvy} for the special case $\varrho_0(t)=1$ and
generalised for any H\"older-continuous measure with spectral
dimension $D<6$ in \cite{Grosse:2019jnv}. First, we need the following
implicitly defined construction:
\begin{definition}\label{defR}
  Let $D\in \{0,2,4\}$ (otherwise take $2\lfloor \frac{D}{2}\rfloor$)
  and $\mu^2>0$. Define implicitly the complex function $R_D(z)$ and
  the deformed measure $\varrho_\lambda(t)$ by the system of equations
\begin{align*}
  R_D(z)=z-\lambda (-z)^{\frac{D}{2}}\int_{\nu_D}^{\Lambda_D^2}
  \frac{dt\,\varrho_\lambda(t)}{(t+\mu^2)^{\frac{D}{2}}(t+\mu^2+z)}\qquad
  \text{and}\qquad 
\varrho_\lambda(t)=\varrho_0(R_D(t)),
\end{align*}
where $R_D(\Lambda^2_D)=\Lambda^2$ and $R_D(\nu_D)=0$. The limit
$\Lambda^2\to \infty$ converges.
\end{definition}
\noindent
This implicitly defined system of equations is for general $\varrho_0$
not exactly solvable. We will give in Sec. \ref{sec.Moyal} two examples
corresponding to the 2- and 4-dimensional Moyal space, where $R_D$ and
$\varrho_\lambda$ can be found explicitly.

Nevertheless, a formal expansion in $\lambda$ of $R_D$ and
$\varrho_\lambda$ can be achieved recursively already in the general
case, starting with
\begin{align}
  \varrho_\lambda(t)&=\varrho_0(t)+\mathcal{O}(\lambda),\quad
  \nonumber
  \\
  R_D(z)&=z-\lambda (-z)^{\frac{D}{2}}\int_{0}^{\Lambda^2}\frac{dt\,\varrho_0(t)}{(t+\overline{\mu}^2)^{\frac{D}{2}}(t+\overline{\mu}^2+z)}+\mathcal{O}(\lambda^2),
  \label{1stroh}
\end{align}
where $\mu^2=\overline{\mu}^2+\mathcal{O}(\lambda)$. The expression is
convergent for $\Lambda^2\to\infty$.

One can prove that the complex function $R_D(z)$ is biholomorphic from
the right half plane
$\{z\in\mathbb{C}:\mathrm{Re}(z)>-\frac{\mu^2}{2}\}$ onto a domain
$U_D\subset\mathbb{C}$ containing $[0,\infty)$ for real $\lambda$
\cite{Grosse:2019jnv}. We define on this domain $U_D$ the inverse
$R^{-1}_D$ (which is not globally defined on $\mathbb{C}$). Then, the
solution of the angle function is obtained by
\begin{theorem}[\cite{Grosse:2019jnv}]\label{Thm:tausolve}
  Let $I_D:U_D\setminus[0,\Lambda^2]\ni w\mapsto I_D(w)\in \mathbb{C}$
  be defined by
\begin{align*}
I_D(w):=-R_D(-\mu^2-R^{-1}_D(w)).
\end{align*}
The consistency relation \eqref{selfcont} is solved by
\begin{align*}
  \tau_a(p)=\lim_{\epsilon\to 0}\mathrm{Im}
  \big(\log(a+I_D(p+\mathrm{i}\epsilon))\big),
\end{align*}
where $\mu^2_{bare}(\Lambda^2)$ is related to $\mu^2$ by
\begin{align*}
	&D=0:&& \mu^2_{bare}(\Lambda^2)=\mu^2\;,\\
		&D=2:&& \mu^2_{bare}(\Lambda^2)=\mu^2-2\lambda\int_{R_D^{-1}(0)}^{R_D^{-1}(\Lambda^2)}\frac{dt\,\varrho_\lambda(t)}{t+\mu^2}\;,\\
		&D=4:&& \mu^2_{bare}(\Lambda^2)=\mu^2-\lambda\mu^2\int_{R_D^{-1}(0)}^{R_D^{-1}(\Lambda^2)}\frac{dt\,\varrho_\lambda(t)}{(t+\mu^2)^2}-2\lambda\int_{R_D^{-1}(0)}^{R_D^{-1}(\Lambda^2)}\frac{dt\,\varrho_\lambda(t)}{t+\mu^2}\;.
\end{align*}
The angle function $\tau_a(p)$ converges in the limit $\Lambda^2\to\infty$.
\end{theorem}\noindent
Equivalently to Theorem \ref{Thm:tausolve} is the statement that the expression in the square brackets of \eqref{eq:GcomplexCont} is equal to $I_D$, that is 
\begin{align*}
	\zeta+\mu^2_{bare}+\lambda\int_0^{\Lambda^2}dt\,\varrho_0(t)\bigg(ZG^{(0)}(\zeta,t)+\frac{1}{t-\zeta}\bigg)=I_D(\zeta)=-R_D(-\mu^2-R^{-1}_D(\zeta)),
\end{align*}
for $\zeta\in U_D\setminus[0,\Lambda^2]$.  The proof of the theorem is
essentially achieved by inserting the solution of $\tau_a(p)$ into rhs
of the consistency relation \eqref{selfcont}, using the system of
implicitly defined functions from Definition \ref{defR} and applying
the Lagrange-B\"urmann inversion theorem, a generalisation of Lagrange
inversion theorem. We refer to \cite{Grosse:2019jnv} for details.

In conclusion, Theorem \ref{Thm:tausolve} together with Definition
\ref{defR} provide the solution of the initial topology $(g,n)=(0,1)$,
generalising the first part of Theorem \ref{throm1} to higher
dimensions. The second part of Theorem \ref{throm1},
i.e.\ the explicit expression
for the 2-point correlation function, extends as follows to $N\to \infty$:
\begin{theorem}[\cite{Grosse:2019jnv}]\label{Thm:2P}
The renormalised 2-point function of the $\lambda\phi^4$ matricial QFT-model
in $D$ dimensions is given by
\begin{align*}
G(a,b):=\frac{\mu^{2 \delta_{4,D}} \exp (N_D(a,b))}{\mu^2+a+b}
\end{align*}
where
\begin{align*}
N_D(a,b)=\frac{1}{2\pi \mathrm{i}}\int_{-\infty}^\infty& dt\bigg\{\log\big(a-R_D(-\tfrac{\mu^2}{2}-\mathrm{i}t)\big)\frac{d}{dt}\log\big(b-R_D(-\tfrac{\mu^2}{2}+\mathrm{i}t)\big)\\
		&-\log\big(a-(-\tfrac{\mu^2}{2}-\mathrm{i}t)\big)\frac{d}{dt}\log\big(b-(-\tfrac{\mu^2}{2}+\mathrm{i}t)\big)\\
		&-\delta_{4,D}\log\big(-R_D(-\tfrac{\mu^2}{2}-\mathrm{i}t)\big)\frac{d}{dt}\log\big(-R_D(-\tfrac{\mu^2}{2}+\mathrm{i}t)\big)\\
		&+\delta_{4,D}\log\big(-(-\tfrac{\mu^2}{2}-\mathrm{i}t)\big)\frac{d}{dt}\log\big(-(-\tfrac{\mu^2}{2}+\mathrm{i}t)\big)
		\bigg\}
\end{align*}
and $R_D$ built by Definition \ref{defR}. For $D\in \{0,2\}$ and for
some restricted cases in $D=4$ (including the $D=4$ Moyal space),
there is an alternative representation
\begin{align}\label{2Pasy}
  &G(a,b)
  \\
  &=\frac{(\mu^2{+}a{+}b)\exp\Bigg\{
    \mbox{\small$\displaystyle\frac{1}{2\pi \mathrm{i}} \int_{-\infty}^\infty dt \log\bigg(\frac{a-R_D(-\frac{\mu^2}{2}{-}\mathrm{i} t)}{a-(-\frac{\mu^2}{2}{-}\mathrm{i} t)}\bigg)\frac{d}{dt}\log\bigg(\frac{b-R_D(-\frac{\mu^2}{2}{+}\mathrm{i} t)}{
   b-(-\frac{\mu^2}{2}{+}\mathrm{i} t)}\bigg)$}\Bigg\}}{(\mu^2+b+R_D^{-1}(a))(\mu^2+a+R_D^{-1}(b))}.\nonumber
	\end{align}
\end{theorem} 
\noindent
We emphasise that $G(a,b)$, as a 2-point function, is by definition
symmetric under $a\leftrightarrow b$. This symmetry is revealed
by integration by parts within the exponential in \eqref{2Pasy}.

Theorem \ref{Thm:2P} provides an exact formula for the 2-point
function for an open neighbourhood around $\lambda=0$ such that a
convergent expansion exists. The following example will give the first
order contribution:
\begin{example}\label{ex:2P}
	Let $\mu^2=1$ for convenience. The first order expansion of $\varrho_\lambda$ and $R_D$ was already given in \eqref{1stroh}. Equivalently, we get for the inverse
	\begin{align*}
		R^{-1}(a)=a+\lambda(-a)^{\frac{D}{2}}\int_{0}^{\Lambda^2}\frac{dt\,\varrho_0(t)}{(t+1)^{\frac{D}{2}}(t+1+a)}+\mathcal{O}(\lambda^2).
	\end{align*}
	Expanding \eqref{2Pasy} of Theorem \ref{Thm:2P}, where
	the exponential does not contribute at the first order since each logarithm starts at order $\lambda$, yields the convergent first order expression
	\begin{align*}
		G(a,b)=\frac{1}{1+a+b}-\frac{\lambda}{(1+a+b)^2}\bigg(&(-a)^{\frac{D}{2}}\int_{0}^{\infty}\frac{dt\,\varrho_0(t)}{(t+1)^{\frac{D}{2}}(t+1+a)}\\
		&+(-b)^{\frac{D}{2}}\int_{0}^{\infty}\frac{dt\,\varrho_0(t)}{(t+1)^{\frac{D}{2}}(t+1+b)}\bigg)+\mathcal{O}(\lambda^2).
	\end{align*}
\end{example}
\noindent 
The reader may also compute the second order contribution in
$\lambda$, where an iterated integral occurs with the canonical
measure $dt\, \varrho_0(t)$. For the exponential, the contour of the
integral should be deformed, and the integrand expands similar to the
computation carried out in \cite[Sec. 7]{Panzer:2018tvy}.

The structure of the solution is still very abstract. The next section
will convey more insight into the implicit definitions of
$\varrho_\lambda$ and $R_D$ and their structure. This implicit system
of equations will be solved for two examples, the 2- and 4-dimensional
Moyal space.

\section{Scalar QFT on Moyal space} \label{sec.Moyal}

Given a real skew-symmetric $D\times D$-matrix $\Theta$, the
associative but noncommutative Moyal product between Schwartz
functions $f,g\in \mathcal{S}(\mathbb{R}^D)$ is defined by
\begin{align}
  (f\star g)(x)=\int_{\mathbb{R}^D\times \mathbb{R}^D}
  \frac{dy\,dk}{(2\pi)^D} \;f(x+\tfrac{1}{2}\Theta k) g(x+y)
  e^{\mathrm{i}\langle k,y\rangle}.
\end{align}
It was understood in \cite{Grosse:2003aj,Grosse:2004yu} that the
scalar QFT on Moyal space which arises from the action functional
\begin{align}
S(\phi)
&:=    \int_{\mathbb{R}^D} \frac{dx}{(8\pi)^{D/2}} \Big(
\frac{1}{2}  \phi \star (-\Delta +4\Omega^2 \|\Theta^{-1}x\|^2+\mu^2) \phi
+\frac{\lambda}{4} \phi\star \phi\star\phi \star \phi\Big)(x)
\label{GW-action}
\end{align}
is perturbatively renormalisable in dimensions $D=2$ and $D=4$.  It
was also noticed \cite{Grosse:2004by}
that the renormalisation group (RG) flow of the
effective coupling constant (in $D=4$, lowest $\lambda$-order) is
bounded and that $\Omega=1$ is a RG fixed point.  Further
investigations therefore focused to $\Omega=1$
\cite{Disertori:2006uy} for which the 
RG-flow of the coupling constant in $D=4$ was proved to be bounded
at any order in perturbation theory \cite{Disertori:2006nq}.

Methods developed in the proofs of these results were decisive in the
derivation \cite{Grosse:2009pa, Grosse:2012uv} of non-linear equation
\eqref{eq:Gcomplex} we started with.

\subsection{$D=2$}

In two dimensions we have
$\Theta=\left(\begin{smallmatrix} 0 & \theta
    \\ -\theta & 0\end{smallmatrix}\right)$. We
introduce a family $(f_{kl})_{k,l\in\mathbb{N}}$ of Schwartz functions 
by
\begin{align}
  f_{kl}(x_1,x_2)=2(-1)^k \sqrt{\frac{k!}{l!}}
  (x_1+\mathrm{i}x_2)^{l-k} L^{l-k}_k\Big(\frac{2x_1^2+2x_2^2}{\theta}\Big)
  e^{-\frac{x_1^2+x_2^2}{\theta}},
\end{align}
where $L^\alpha_k$ are associated Laguerre polynomials.
It is straightforward to prove  
\begin{align}
\overline{f_{kl}}&=f_{lk}\;,\qquad
  f_{kl}\star f_{mn}=\delta_{lm} f_{kn}\;,\qquad
  \int_{\mathbb{R}^2} dx\;f_{kl}(x)=2\pi \theta \delta_{kl}\;,
  \nonumber
  \\
&  (-\partial_{x_1}^2-\partial_{x_2}^2+ \frac{4}{\theta^2} (x_1^2+x_2^2))
  f_{kl}(x_1,x_2)= \frac{4}{\theta} (k+l+1) f_{kl}(x_1,x_2).
\end{align}
Therefore, expanding $\phi(x)=\sum_{k,l=0}^\infty \phi_{kl} f_{kl}(x)$, we turn
\eqref{GW-action} for $D=2$ into $S(\phi)=\lim_{N\to \infty} S_N(\phi)$ with
\begin{align}
  S_N(\phi)=\frac{\theta}{4}\Big(
  \sum_{k,l=0}^{N-1}\Big( \frac{4}{\theta}k
  +\frac{\mu^2}{2}+\frac{2}{\theta}\Big)\phi_{kl}\phi_{lk}
+\frac{\lambda}{4} \sum_{k,l,m,n=0}^{N-1}
\phi_{kl}\phi_{lm}\phi_{mn}\phi_{nk}\Big).
\end{align}
Setting here $\frac{\theta}{4}=\frac{N}{\Lambda^2}$ and
$E_k=\Lambda^2 \frac{k}{N}+ \frac{\mu^2}{2}+\frac{\Lambda^2}{2N}$
we recover the Minlos measure (\ref{minlos-q-int}) as 
\begin{align}
  d\mu_{int}(\phi)= \frac{1}{\mathcal{Z}}
  \exp(-\Lambda^2 S_N(\phi)) 
\prod_{k=0}^{N-1}d\phi_{kk}
\prod_{k=1}^{N-1}\prod_{l=0}^k d \mathrm{Re}\phi_{kl} \,
d \mathrm{Im}\phi_{kl}\;.
\end{align}
After some shift by the lowest spectral value $E_0$ which is absorbed in
the bare mass we identifty $e_k=\Lambda^2 \frac{k}{N}$ and $r_k=1$ in the
spectral measure 
$\varrho_0(t)=\frac{1}{N}\sum_{k=0}^{N-1} \delta(t-e_k)$.
For $N\to \infty$ this converges, in the sense of Riemann sums,
to the characteristic function on $[0,\Lambda^2]$. Compared with
Definition \ref{defR} we recover $D=2$ and obtain for the function
$R_2$ in the limit $\Lambda^2\to \infty$
\begin{align}
R_2(z)=z+\lambda z\int_0^\infty\frac{dt}{(t+\mu^2)(t+\mu^2+z)}=z+\lambda \log\Big(1+\frac{z}{\mu^2}\Big).
\end{align}
The function $R_2$ is holomorphic on $\mathbb{C}\setminus]-\infty,-\mu^2]$.
It can be inverted there in terms of the Lambert-$W$ function 
defined by $W(z)e^{W(z)}=z$ to
\begin{align*}
	R^{-1}_2(z)=\lambda W\Big(\frac{\mu^2}{\lambda}e^{\frac{z+\mu^2}{\lambda}}\Big)-\mu^2.
\end{align*}
Similar to the logarithm, the Lambert-$W$ function has infinitely
many branches $W_k$ labelled by $k\in \mathbb{Z}$,
where the inverse $R^{-1}$ is obtained by the principal branch $W_0$.
Following \cite{Panzer:2018tvy}, this inverse admits a convergent expansion 
\begin{align*}
	R^{-1}_2(z)=\sum_{n=1}^\infty \frac{(-\lambda)^n}{n!}\frac{d^{n-1}}{dz^{n-1}}\log\Big(1+\frac{z}{\mu^2}\Big)^n
\end{align*}
consisting only of powers of the logarithm. 

Carrying out higher order computations on the $D=2$ Moyal plane shows
that further transcendental functions occur which are not representable
via powers of logarithms. These functions are generated by the
integral inside the exponential of \eqref{2Pasy}, they are
essentially in the class of Nielsen polylogarithms (see
\cite{Panzer:2018tvy} for more detials).

Applying the construction of Sec. \ref{Sec.BTR}, a completely different structure of the spectral curve is revealed, which is built of
\begin{align*}
	x(z)=z+\lambda \log\Big(1+\frac{z}{\mu^2}\Big),\qquad y(z)=-x(-z-\mu^2)=z+\mu^2-\lambda\log\Big(-\frac{z}{\mu^2}\Big).
\end{align*}
This curve is no longer of algebraic nature. The continuum limit from
an algebraic curve (as in \eqref{specurve}) to a transcendental
spectral curve is still rather mysterious. The number of branches of
$R^{-1}$ tends in the limit to infinity, where the number of
ramification points increases as well. However, these ramification
points accumulate in the continuum to a single ramification point $\beta$
at
\begin{align*}
  x'(\beta)=R'_2(\beta)=1+\frac{\lambda}{\mu^2+\beta}=0
  \quad\Rightarrow\quad  \beta=-\mu^2-\lambda.
\end{align*}
In the limit $\lambda\to 0$, the ramification point $\beta$ approaches
the singularity of the logarithm, where the full construction becomes
meaningless.

However for positive $\lambda$, we may ask whether the model is still
governed by the universal structure of Theorem \ref{thmBTRplan}. This
would imply rationality for all $\Omega^{(0)}_n(z_1,...,z_n)$ in
$z_i$. Inserting $z_i=R_2^{-1}(x_i)$, we could verify that the
$\Omega^{(0)}_n$ are expanded only into powers of logarithms, and
therefore remain in a simple class of functions. Whereas,
correlation functions in general are generated by generalisations of
polylogarithms, like Nielsen polylogarithms.

\begin{figure}[h!]
  \centering
  \includegraphics[width= 0.99\textwidth]{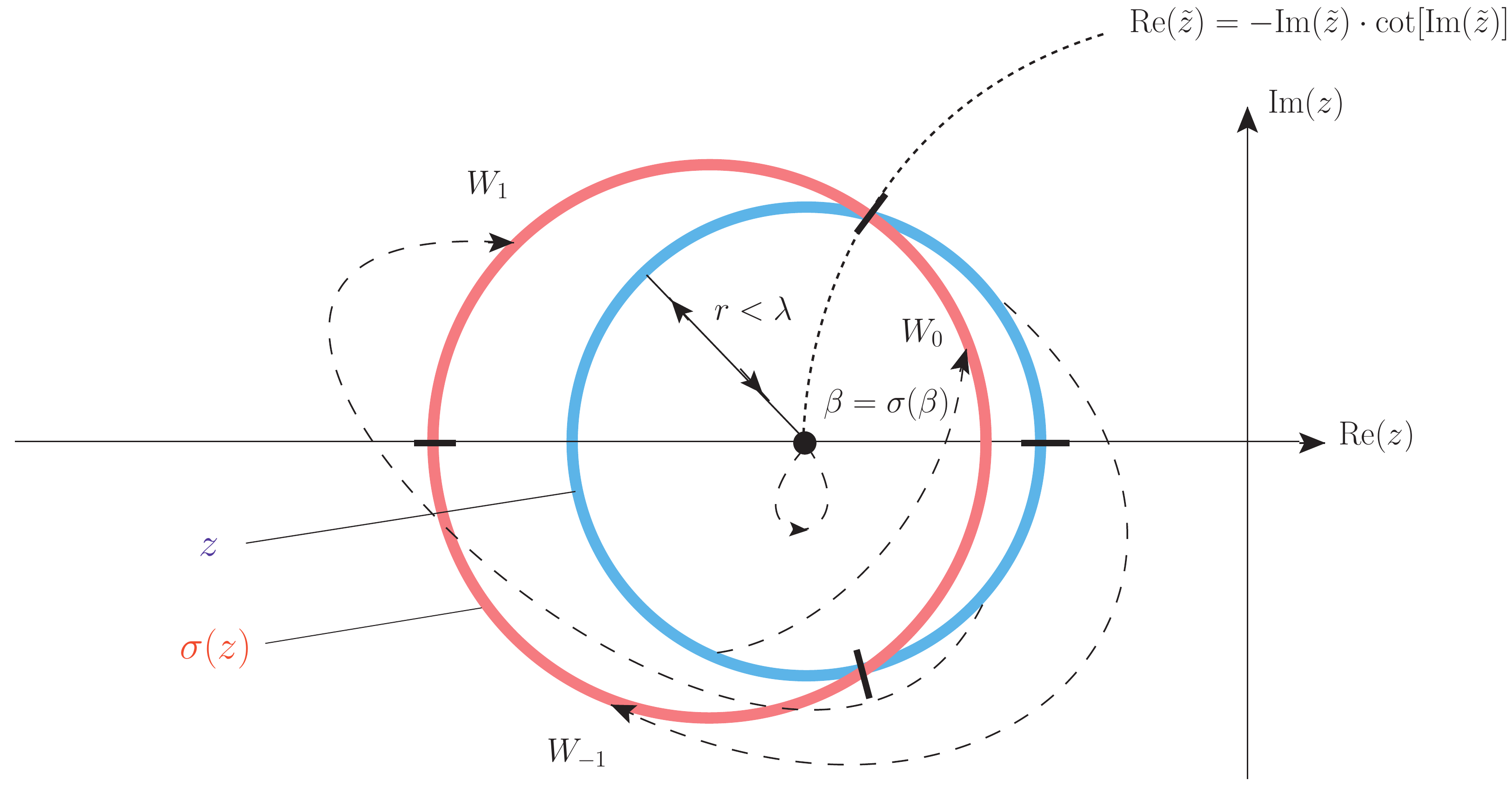} 
    \caption{Symbolical interpretation of the local Galois involution $\sigma(z)$ (red) which can be defined piecewise using the three branches of the Lambert $W$-function $W_{0,\pm 1}$. The definition as Galois involution holds for $ |z-\beta|<\lambda$ giving a concrete meaning to the term \textit{local} involution.  \label{galois2d}}
\end{figure}

The local Galois involution becomes fairly easy to handle:
\begin{lemma}
  Let
  $\mathcal{V}:= \{z\in \mathbb{C}\;|~|z-\beta| <\lambda\}$
  and  $\tilde z(z)=\frac{\exp[(1+z)/\lambda](1+z)}{\lambda}$ with
  $\tilde z(\beta)=-1/e$, the threefold branch point.  Then the local
  Galois involution $\sigma_{2}(z) $ with $R_2(z)=R_2(\sigma_{2}(z))$
  and fixed point $\sigma_2(\beta)=\beta$ is piecewise defined within
  three branches of the Lambert W-function as
\begin{align*}
&\sigma_{2}(z) = -1+\lambda W_k(\tilde z)
\\
&\text{with }
\begin{cases}
                    k=-1 & \text{if $\{\mathrm{Re}(\tilde z)\geq -1/e ,|\mathrm{Im}(\tilde z)| \cot [|\mathrm{Im}(\tilde z)|] \leq \mathrm{Re}(\tilde z) \}$}\\
                    k=1 & \text{c.c. of $k=-1$ sector } \\
                    k=0 & \text{else}
                    \end{cases}
\end{align*}
\end{lemma}  
Figure~\ref{galois2d} gives a picture of this local Galois involution.

\subsection{$D=4$}
By transform of variables one can 
achieve $\Theta=\left(\begin{smallmatrix} 0 & \theta
    \\ -\theta & 0\end{smallmatrix}\right)\otimes \mathbb{I}_{2\times 2}$.
We expand $\phi(x_1,x_2,x_3,x_4)=\phi_{\substack{k_1l_1\\k_2l_2}}
f_{k_1l_1}(x_1,x_2)f_{k_2l_2}(x_3,x_4)$.
Using the Cantor bijection
$P\binom{k_1}{k_2} := k_2+\frac{1}{2}(k_1+k_2 )(k_1+k_2+1)$
between $\mathbb{N}^2$ and $\mathbb{N}$ we can map
$\phi_{\substack{k_1l_1\\k_2l_2}}$ to standard matrix elements
$\phi_{kl}$. Setting $|k|:=k_1+k_2$ if $P\binom{k_1}{k_2}=k$, 
the action functional \eqref{GW-action} takes for $D=4$ 
the form $S(\phi)=\lim_{N\to \infty} S_N(\phi)$ with
\begin{align*}
  S_N(\phi)=\Big(\frac{\theta}{4}\Big)^2\Bigg(&
  \sum_{\substack{k,l=0 \\ |k|,|l|< \sqrt{N}}}^\infty
\frac{4}{\theta} |k|
+\frac{\mu^2}{2}+\frac{4}{\theta}\Big)
\phi_{kl}\phi_{lk}
+\frac{\lambda}{4}   \sum_{\substack{k,l,m,n=0 \\
    |k|,|l|,|m|,|n|< \sqrt{N}}}^\infty
\hspace*{-1ex} 
\phi_{kl}\phi_{lm}\phi_{mn}\phi_{nk}\Big)
\Bigg).
\end{align*}
There are $n+1$ natural numbers $k$ with $|k|=n$
and $\frac{(n+1)(n+2)}{2}$
natural numbers $k$ with $|k|\leq n$.
Setting $\frac{\theta}{4}=\frac{\sqrt{N}}{\Lambda^2}$ and
$e_n=\Lambda^2 \frac{n}{\sqrt{N}}+ \frac{\mu^2}{2}+\frac{\Lambda^2}{\sqrt{N}}$
we recover the Minlos measure (\ref{minlos-q-int}) as 
\begin{align}
  d\mu_{int}(\phi)= \frac{1}{\mathcal{Z}}
  \exp(-\Lambda^4 S_N(\phi)) d\phi
\end{align}
where the spectral values $e_n$ have multiplicity $r_n=n+1$. 
After some shift by the lowest spectral value $e_0$, which is absorbed in
the bare mass, the 
spectral measure 
$\varrho_0(t)=\frac{1}{N}\sum_{n=0}^{\sqrt{N}-1} r_n \delta(t-e_n)$
converges for $N\to \infty$, in the sense of Riemann sums,
to $t\chi_{[0,\Lambda^2]}(t)$, where $\chi_{[0,\Lambda^2]}$ is
the characteristic function of $[0,\Lambda^2]$. 
Compared with Definition \ref{defR} we recover $D=4$ and
obtain the function $R_4$ in the limit $\Lambda^2\to \infty$
as
\begin{align}\label{linInt}
  R_4(z)=z-\lambda z^2\int_0^\infty
  \frac{dt\, R_4(t)}{(t+\mu^2)^2(t+\mu^2+z)}.
\end{align}
The leading order expansion is easily computed
\begin{align*}
	R_4(z)=z+\lambda \bigg( z-\mu^2\Big(1+\frac{z}{\mu^2}\Big)\log\Big(1+\frac{z}{\mu^2}\Big)\bigg)+\mathcal{O}(\lambda^2).
\end{align*}
From the expansion, one would expect that $R_4(z)$ has linear
asymptotics for $z$ tending to infinity. However, this is not the
case: The asymptotic behaviour is only visible by the resummation of
all orders in $\lambda$, which is given in terms of a Gau\ss{}ian
hypergeometric function:
\begin{theorem}[\cite{Grosse:2019qps}]
	The linear integral equation \eqref{linInt} is solved by
	\begin{align}
		R(z)=z  \;_2F_1\Big(\genfrac{}{}{0pt}{}{
			\alpha_\lambda,\;1-\alpha_\lambda}{2}\Big|-\frac{z}{\mu^2}\Big),\quad
		\text{where } ~
		\alpha_\lambda:=\begin{cases}
                  \frac{\arcsin(\lambda\pi)}{\pi} & \text{for }|\lambda|\leq
                  \frac{1}{\pi} \;,\\ \frac{1}{2}
                  +\mathrm{i} \frac{\mathrm{arcosh}(\lambda\pi)}{\pi}
                  & \text{for } \lambda \geq \frac{1}{\pi}\;.
                \end{cases}
              \end{align}
\end{theorem}\noindent
The proof is obtained by inserting the result into the linear integral
equation and using identities of the Gaussian hypergeometric function.

The important fact is that the linear dependence of $\lambda$ within
the integral equation \eqref{linInt} is packed into a highly nonlinear
dependence given by the $\arcsin$-function into the coefficients of
the hypergeometric function. Making use of that, a different
asymptotic behaviour between $\varrho_\lambda$ and $\varrho_0$ is
observed. It is well-known that the hypergeometric function behaves like
\begin{align*}
	\,_2F_1\Big(\genfrac{}{}{0pt}{}{
          a,\;1-a}{2}\Big|-x\Big) \stackrel{x\to \infty}{\sim}
        \frac{1}{x^a}\quad \text{for } |a|<\frac{1}{2}.
\end{align*} 

Together with the definition of the spectral dimension
\eqref{specdim}, we conclude:
\begin{corollary}[\cite{Grosse:2019qps}]
  For $|\lambda|<\frac{1}{\pi}$, the deformed measure
  $\varrho_\lambda=R_4$ of four-dimensional Moyal space has the
  spectral dimension $D=4-2\frac{\arcsin(\lambda \pi)}{\pi}$.
\end{corollary}\noindent
The theory admits on the four-dimensional Moyal space a dimensional
drop to an effective spectral dimension related to an effective
spectral measure. This is revealed by knowledge of the exact
solutions.

From a quantum field theoretical perspective, this dimension drop is the
most important result. It means \emph{that the $\lambda\phi^{*4}$-QFT
  model on 4D Moyal space is non-trivial}, i.e.\ consistent over any
(energy) scale $\Lambda^2\to \infty$. If we had $R_4(z)\propto z$ as
expected from a perturbative expansion then the integral in
\eqref{linInt} had to be restricted to $[0,\Lambda^2]$ (otherwise
$R^{-1}$ would be not defined on $\mathbb{R}_+$). We recall that
the Landau ghost problem \cite{Landau:1954??}, or triviality, is a
severe threat to quantum field theory. It almost killed
renormalisation theory in the 1960s, rescued then by discovery of asymptotic
freedom rescued non-Abelian Yang-Mills theories. But these theories
are complicated; they require a non-perturbative treatment to deal
with confinement. A simple 4D QFT-model without triviality problem was
not known so far. For the $\lambda\phi^4$-model, triviality was proved
in $D=4+\epsilon$ dimensions \cite{Aizenman:1981du, Frohlich:1982tw}
in the 1980s. (Non-)triviality in $D=4$ dimensions remained an open
problem for almost 40 years; only recently Aizenman and Duminil-Copin
achieved the proof \cite{Aizenman:2019yuo} of (marginal) triviality.
Therefore, the construction of a simple, solvable and non-trivial
QFT-model in four dimensions (albeit on a noncommutative space) is a
major achievement for renormalisation theory.

In the next step one aims to apply the construction of (B)TR as explained in
Sec.~\ref{Sec.BTR}. But here we are faced with a problem: Defining
$x(z)=R_4(z)$, the derivative $x'(z)$ is no longer in the class of
rational functions, which was a fundamental assumption in (B)TR. In other words,
the full construction of (B)TR based on algebraic, not transcendental,
ramification points fails directly from the beginning. It remains to
investigate in a long-term project whether an adapted formulation of  
(B)TR can extend the previous results to four dimensions.

\section{Cross-checks: perturbative expansions}

In nowadays particle physics, one is often dependent on
approximative methods like expansions of correlation functions into
series of Feynman graphs. A main advantage of investigating the
present toy model is undoubtedly the fact that we only use the Feynman
graph expansion to perform a cross-check of our exact solutions.
Due to its nature as a matrix model, the expansion of the cumulants
creates \textit{ribbon graphs} (also \textit{fat Feynman graphs}). By
their duality to maps on surfaces, we also have points of contact
with enumerative geometry and combinatorics.

\subsection{Ribbon graphs}

In order to obtain a perturbative series of the cumulants
(\ref{partition}), one expands the exponential
$\exp(-\frac{\lambda N}{4} \mathrm{Tr}(\phi^4))$ inside the measure
(\ref{minlos-q-int}) into a (formal) power series in the
\textit{coupling constant} $\lambda$.  As result we obtain products of
matrix elements $\phi_{kl}$ from the expansion
$(\mathrm{Tr}(\phi^4))^v$ of the exponential and from the products
present in (\ref{partition}), integrated against the Gau\ss{} measure
(\ref{minlos-q}). The resulting sum over pairings can be interpreted
as ribbon graphs. A ribbon has two strands, which carry a
labelling. Two strands connected by a four-valent vertex are
identified. We denote by $\mathfrak{G}^{v,\pi}_{k_1,...,k_n}$ the set
of labelled connected ribbon graphs with $v$ four-valent vertices and
$n$ one-valent vertices, where the strands connected to the one-valent
vertices (from the $n$ factors in (\ref{partition})) are labelled by
$(k_1,k_{\pi(1)})$, \dots, $(k_n,k_{\pi(n)})$. It is assumed here that
the $k_i$ in (\ref{partition}) are pairwise different so that
$l_i=\pi(k_i)$ for a permutation
$\pi in \mathcal{S}_{n}$.
Let a graph $\Gamma$
have $r=2v+n/2$ ribbons and $s(\Gamma)$ loops (closed strands). We
associate a weight $\varpi(\Gamma)$ to $\Gamma$ by applying the
following \textit{Feynman rules}:
\begin{itemize}\itemsep -1pt
\item every 4-valent ribbon-vertex carries  a factor $-\lambda$ 
\item every ribbon with strands labelled by $p,q$ carries a factor $\frac{1}{e_p+e_q}$ (\textit{propagator})
\item multiply all factors and apply the summation operator
  $\frac{1}{N^s}\sum_{l_1,..,l_s=1}^N r_{l_1}...r_{l_s}$ over the $s=s(\Gamma)$ loops
  (closed strands) labelled by $l_1,...,l_s$.
\end{itemize}
Then the cumulants expand for  pairwise different $k_1,...,k_n$
and
$n$ even into the following series:
\begin{align}
 \sum_{v=0}^\infty \sum_{\Gamma \in \mathfrak{G}^{v,\pi}_{k_1,...,k_n}}
  N^{v-r+n+s(\Gamma)} \varpi(\Gamma).
\label{expansion}
\end{align}
The cyclic order of the ribbon vertex gives an orientation of the
full ribbon graph such that we achieve a natural embedding of the
graph into a Riemann surface. The exponent of $N^{v-r+n+s(\Gamma)}$
represents the Euler characteristic $\chi=2g-2+b$ of the Riemann
surface with $b$ the number of boundary components and $g$ the
genus. Furthermore, the permutation $\pi$ splits $(k_1,k_{\pi(1)})$,
\dots, $(k_n,k_{\pi(n)})$ into blocks with cyclic symmetry,
see \eqref{eq:cumulants}. This suggests the notation
$G_{|k_1^1\dots k_{n_1}^1|\dots |k_1^b\dots k_{n_b}^b|}$
where the blocks are seperated by the vertical lines and each
block is cyclically symmetric (see \cite{Branahl:2020uxs} for more details).
We will denote
the set of these ribbon graphs by
$\mathfrak{G}^{v}_{|k_1^1...k_{n_1}^1|...|k_1^b...k_{n_b}^b|}
=\bigcup_{g=0}^\infty
\mathfrak{G}^{g,v}_{|k_1^1...k_{n_1}^1|...|k_1^b...k_{n_b}^b|}$, so
that finally the expansion reads:\enlargethispage{5mm}
\begin{align}
G^{(g)}_{|k_1^1...k_{n_1}^1|...|k_1^b...k_{n_b}^b|}
=\sum_{v=0}^\infty \sum_{\Gamma \in
  \mathfrak{G}^{g,v}_{|k_1^1...k_{n_1}^1|...|k_1^b...k_{n_b}^b|}}
  \varpi(\Gamma)\;.
\label{cumulants-3}
\end{align}

\subsection{Perturbative renormalisation of ribbon graphs}

As mentioned before, a QFT is achieved in a continuum limit
$N\to \infty$. In sec.~\ref{sec.Moyal} we have implemented this
limit in two steps. We first passed to continuous spectral measure
$\varrho_0$ on a finite interval $[0,\Lambda^2]$ and then arrived at a
QFT in the limit $\Lambda^2\to \infty$.  This procedure gives rise to
slightly different Feynman rules. The summations converge to
integrals, which are not necessarily convergent (by themselves) after
removing the cut-off $\Lambda^2$. The perturbative renormalisation to
avoid divergencies is well understood and of similar structure as
renormalisation in local QFT. The overlapping divergencies are
controlled by Zimmermann's forest formula (see
e.g. \cite{Hock:2020rje}). In the work \cite{Thurigen:2021zwf} even
more is proved. It is known that certain graphs (so-called 2-graphs,
where ribbon graphs are a special case) have a Hopf-algebraic
structure. The application of the antipode of the Hopf algebra proves
Zimmermann's forest formula for the 2-graphs in general
\cite{Thurigen:2021ntr}. This is analogous to the techniques of
Connes and Kreimer \cite{Connes:1999yr}.

We refer to these works and references therein for more details, but
want to present a rather trivial example, where the forest consists only
of the empty forest.
\begin{example}
The two graphs of the 2-point function of order $\lambda$ are given in
Fig. \ref{fig:2PGraphs}.
\begin{figure}[h!]
	\centering
	\def\svgwidth{400pt}    
\begingroup%
  \makeatletter%
  \providecommand\color[2][]{%
    \errmessage{(Inkscape) Color is used for the text in Inkscape, but the package 'color.sty' is not loaded}%
    \renewcommand\color[2][]{}%
  }%
  \providecommand\transparent[1]{%
    \errmessage{(Inkscape) Transparency is used (non-zero) for the text in Inkscape, but the package 'transparent.sty' is not loaded}%
    \renewcommand\transparent[1]{}%
  }%
  \providecommand\rotatebox[2]{#2}%
  \newcommand*\fsize{\dimexpr\f@size pt\relax}%
  \newcommand*\lineheight[1]{\fontsize{\fsize}{#1\fsize}\selectfont}%
  \ifx\svgwidth\undefined%
    \setlength{\unitlength}{446.93579482bp}%
    \ifx\svgscale\undefined%
      \relax%
    \else%
      \setlength{\unitlength}{\unitlength * \real{\svgscale}}%
    \fi%
  \else%
    \setlength{\unitlength}{\svgwidth}%
  \fi%
  \global\let\svgwidth\undefined%
  \global\let\svgscale\undefined%
  \makeatother%
  \begin{picture}(1,0.40734917)%
    \lineheight{1}%
    \setlength\tabcolsep{0pt}%
    \put(0,0){\includegraphics[width=\unitlength,page=1]{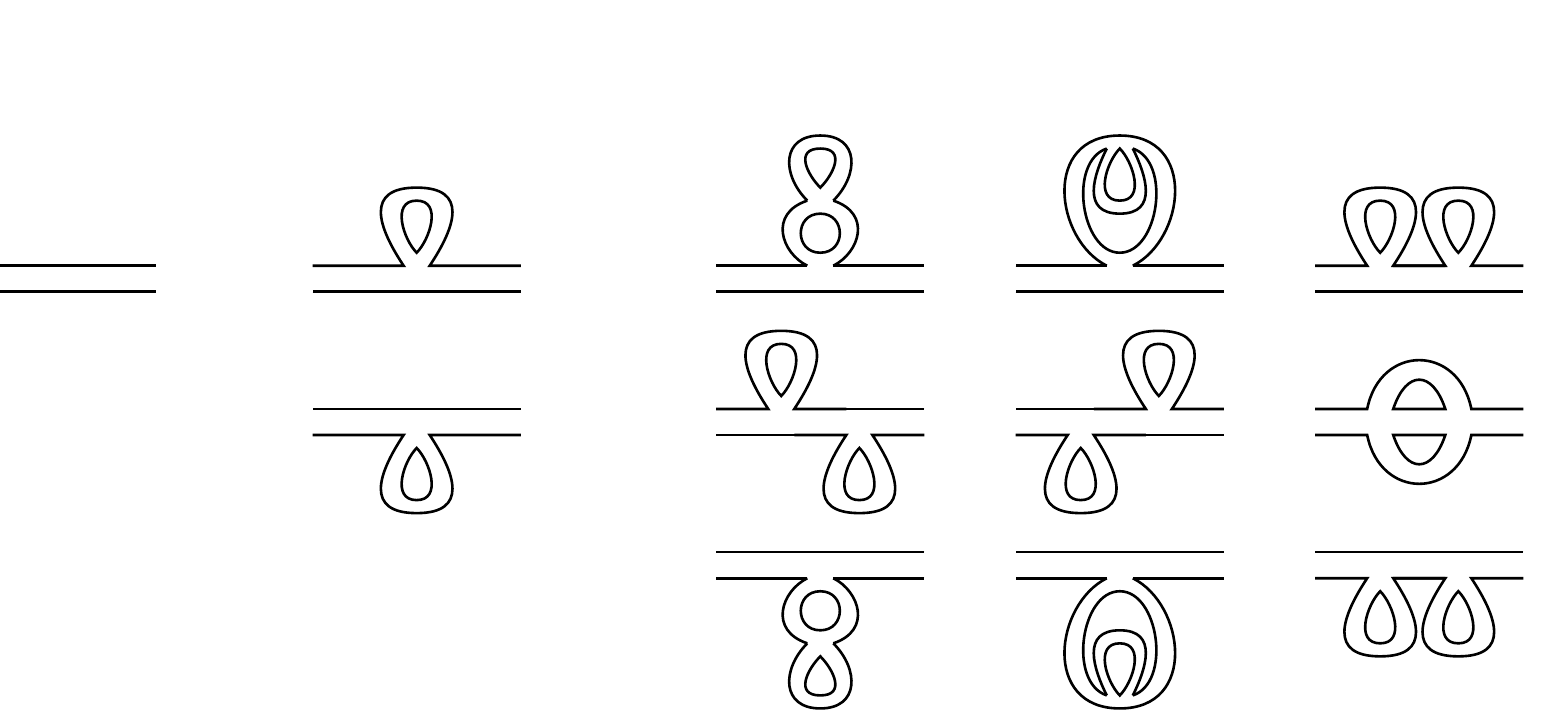}}%
    \put(0.02308892,0.39271376){\makebox(0,0)[lt]{\lineheight{1.25}\smash{\begin{tabular}[t]{l}$\lambda^0$\end{tabular}}}}%
    \put(0.25979371,0.39171405){\makebox(0,0)[lt]{\lineheight{1.25}\smash{\begin{tabular}[t]{l}$\lambda^1$\end{tabular}}}}%
    \put(0.69792114,0.39335028){\makebox(0,0)[lt]{\lineheight{1.25}\smash{\begin{tabular}[t]{l}$\lambda^2$\end{tabular}}}}%
  \end{picture}%
\endgroup%

	\caption{All graphs contributing to the 2-point function $G^{(0)}_{|ab|}$ up to order $\lambda^2$, where the upper strand is labelled by $a$ and the lower by $b$ for each graph. Topologically,
		some graphs are the same but different elements of $\mathfrak{G}^{0,v}_{|ab|}$
		due to different labellings.}    
	\label{fig:2PGraphs}          
\end{figure}
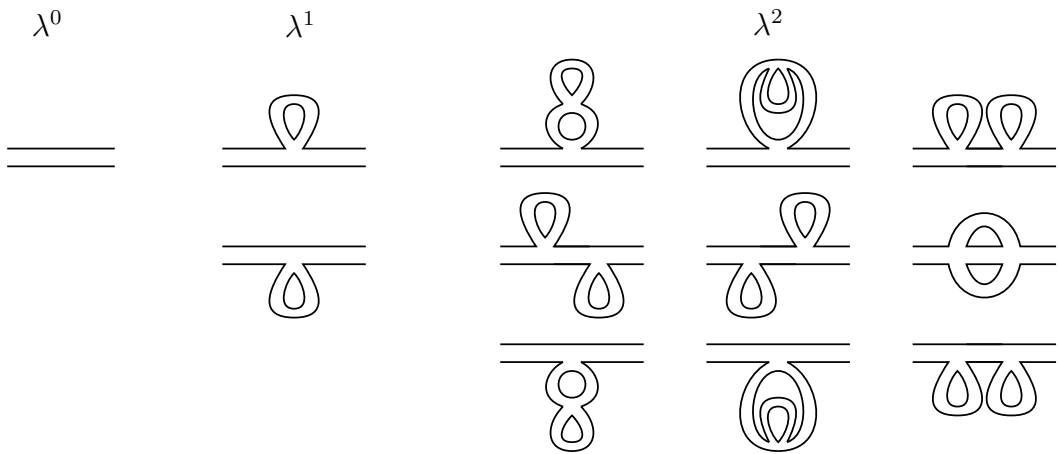
The forest formula is trivial and
realised by a Taylor subtraction depending on the dimension of the
theory given by the spectral measure $\varrho_0(t) dt$. In $D$
dimensions this measure behaves asymptotically as
$\varrho_0(t) dt\sim t^{\frac{D}{2}-1}dt$. For the first graph of
order $\lambda$ with external labelling $a,b$, we get (put $\mu^2=1$
for convenience)
	\begin{align*}
		&\frac{-\lambda}{(1+a+b)^2}\int_0^\infty \varrho(t)dt\bigg(\frac{1}{1+a+t}-T^{D/2-1}_a\Big(\frac{1}{1+a+t}\Big)\bigg)\\
		=&\frac{-\lambda}{(1+a+b)^2}\int_0^\infty \varrho(t)dt\bigg(\frac{1}{1+a+t}-
		\frac{1}{1+t}...-\frac{a^{\frac{D}{2}-1}}{(\frac{D}{2}-1)!}\frac{d^{\frac{D}{2}-1}}{dw^{\frac{D}{2}-1}}\frac{1}{1+t+w}\vert_{w=0}\bigg)\\
		=&\frac{-\lambda}{(1+a+b)^2}\int_0^\infty \varrho(t)dt\frac{(-a)^{\frac{D}{2}}}{(1+a+t)(1+t)^{\frac{D}{2}}}.
	\end{align*}
	Taking the second graph of Fig. \ref{fig:2PGraphs} into account, we have to interchange $a$ with $b$ only. Adding both expressions, we confirm the result of Example \ref{ex:2P} for any $D$, which was the expansion of the explicit result. 
\end{example}
We refer to \cite{Hock:2020rje} for further explicit examples. Already
the second order of the 2-point function involves tricky computations,
since the full beauty of Zimmermann's forest formula is required for
the sunrise graph at $D=4$. An interested reader could try this
computation and compare it with the expansion of the exact solution of
Theorem \ref{Thm:2P}, where a contour integral around the branch cut
of a sectionally holomorphic function has to be evaluated.

\subsection{Boundary creation}

In this subsection we will focus on the perturbative interpretation of
the objects constructed in Sec.~\ref{Sec.BTR}.  The objects
$\Omega_{g,n}$ form the fundamental building blocks of the theory. We
will forget about the continuum limit and work in the discrete setting
again, since their construction in $D=4$ is not yet understood.

It was shown in \cite{Branahl:2020uxs} that the
$\Omega^{(g)}_{q_1,...,q_m}$ are special polynomials of the cumulants
$G^{(g)}_{\dots}$. For $m=1$ this is the definition
$\Omega^{(g)}_{q} := \frac{1}{N}\sum_{k=1}^N
G^{(g)}_{|qk|}+\frac{1}{N^2}G^{(g)}_{|q|q|}$
given after (\ref{eq:Omega-gm}). For $m=2$ we have:
\begin{proposition}[\cite{Branahl:2020uxs}]
  \label{prop:Om02G}
  For $q_1\neq q_2$ one has
  \begin{align}
\Omega^{(g)}_{q_1,q_2}   &=
\frac{\delta_{g,0}}{(e_{q_1}-e_{q_2})^2}
+\sum_{g_1+g_2=g} G^{(g_1)}_{|q_1q_2|} G^{(g_2)}_{|q_1q_2|}
\nonumber
\\
&
+    \frac{1}{N^2}\sum_{k,l=1}^N G^{(g)}_{|q_1k|q_2l|}
+\frac{1}{N}\sum_{k=1}^N \Big(G^{(g)}_{|q_1kq_1q_2|}+G^{(g)}_{|q_2kq_2q_1|}
+G^{(g)}_{|q_1kq_2k|}\Big)
\nonumber
\\
& +\frac{1}{N} \sum_{k=1}^N \Big(G^{(g-1)}_{|q_1k|q_2|q_2|}
+G^{(g-1)}_{|q_2k|q_1|q_1|}\Big)
+G^{(g-1)}_{|q_1q_2q_2|q_2|}+G^{(g-1)}_{|q_2q_1q_1|q_1|}
\nonumber
\\
&+\sum_{g_1+g_2=g-1} G^{(g_1)}_{|q_1|q_2|} G^{(g_2)}_{|q_1|q_2|}
+G^{(g-2)}_{|q_1|q_1|q_2|q_2|}\;.
\end{align}  
\end{proposition}
Inserting the ribbon graph expansions of the cumulants $G^{(g)}$ one
obtains in this way a ribbon graph expansion of the
$\Omega^{(g)}_{q_1,...,q_m}$.

There is an alternative way to produce these ribbon graph expansions.
We recall from (\ref{eq:Omega-gm}) that the
$\Omega^{(g)}_{q_1,...,q_m}$ are defined as derivatives of
$\Omega_{q_1}^{(g)}$ with respect to the spectral values $e_{q_i}$.
We declare this derivative as a \textit{boundary creation operator}
$\hat{T}_q:=-N \frac{\partial}{\partial e_q}$
(looking at its action, its name is self-explanatory). One
can go one step further and think about a primitive of $\Omega^{(g)}_q$
under the creation operator,
$\Omega_q^{(g)}=:\hat{T}_q \mathcal{F}^{(g)}$.
The free energies $\mathcal{F}^{(g)}$ defined in this way 
agree with the genus expansion of the  logarithm
of the partition function itself:
\begin{align}
          \log \mathcal{Z}
          =\sum_{g=0}^\infty\sum_{v=0}^\infty 
          \sum_{\Gamma_0\in\mathfrak{G}^{g,v}_\emptyset}
          \frac{N^{2-g}\varpi(\Gamma_0)}{|\mathrm{Aut}(\Gamma_0)|}\;,
\end{align} 
where $\varpi(\Gamma_0)$ is given by the same Feynman rules as
before. This expansion creates closed ribbon graphs without any
boundary (in physics vocabulary: \textit{vacuum graphs}).
In contrast to the previously considered open graphs, 
closed graphs have a non-trivial automorphism group.

\begin{figure}[h!]             
	\centering
	\def\svgwidth{400pt}    
\begingroup%
  \makeatletter%
  \providecommand\color[2][]{%
    \errmessage{(Inkscape) Color is used for the text in Inkscape, but the package 'color.sty' is not loaded}%
    \renewcommand\color[2][]{}%
  }%
  \providecommand\transparent[1]{%
    \errmessage{(Inkscape) Transparency is used (non-zero) for the text in Inkscape, but the package 'transparent.sty' is not loaded}%
    \renewcommand\transparent[1]{}%
  }%
  \providecommand\rotatebox[2]{#2}%
  \newcommand*\fsize{\dimexpr\f@size pt\relax}%
  \newcommand*\lineheight[1]{\fontsize{\fsize}{#1\fsize}\selectfont}%
  \ifx\svgwidth\undefined%
    \setlength{\unitlength}{513.86557695bp}%
    \ifx\svgscale\undefined%
      \relax%
    \else%
      \setlength{\unitlength}{\unitlength * \real{\svgscale}}%
    \fi%
  \else%
    \setlength{\unitlength}{\svgwidth}%
  \fi%
  \global\let\svgwidth\undefined%
  \global\let\svgscale\undefined%
  \makeatother%
  \begin{picture}(1,0.30156411)%
    \lineheight{1}%
    \setlength\tabcolsep{0pt}%
    \put(0.28905551,0.28161671){\makebox(0,0)[lt]{\lineheight{1.25}\smash{\begin{tabular}[t]{l}$\lambda^1$\end{tabular}}}}%
    \put(0.7372662,0.28199349){\makebox(0,0)[lt]{\lineheight{1.25}\smash{\begin{tabular}[t]{l}$\lambda^2$\end{tabular}}}}%
    \put(0,0){\includegraphics[width=\unitlength,page=1]{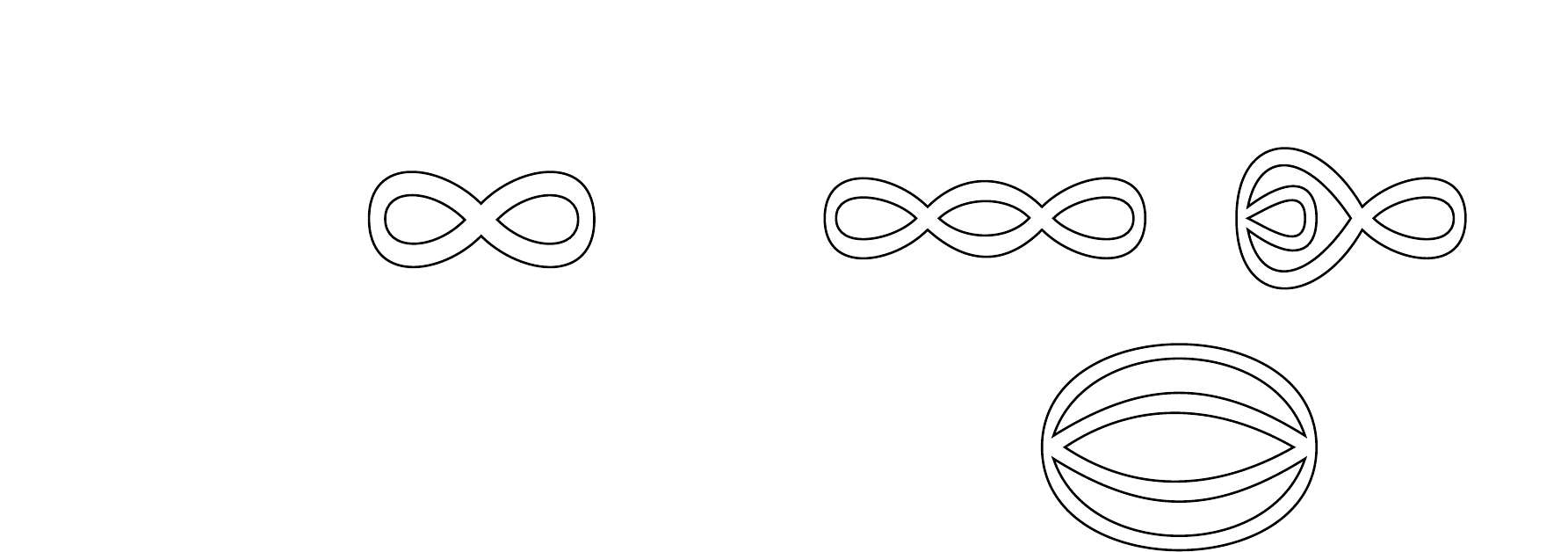}}%
    \put(0.04445554,0.28110333){\makebox(0,0)[lt]{\lineheight{1.25}\smash{\begin{tabular}[t]{l}$\lambda^0$\end{tabular}}}}%
    \put(0,0){\includegraphics[width=\unitlength,page=2]{freeenergy.pdf}}%
  \end{picture}%
\endgroup%
  
	\caption{The graph at order $\lambda^0$ is added as the empty
          ribbon graph. All these graphs contribute to the free energy
          of genus 0 up to order $\lambda^2$. These graphs are
          elements of $\mathfrak{G}^{0,v}_\emptyset$. The melon graph $\Gamma_M$ (in the second row) has
          $ |\mathrm{Aut}(\Gamma_M)|=8$ and the other four graphs
          $ |\mathrm{Aut}(\Gamma)|=2$. } 
	\label{fig:freeenergy}          
\end{figure}

Let us focus on the planar sector to illustrate the structures. We
start with $\mathcal{F}^{(0)}$ and describe geometrically how
the operator $\hat{T}_q$ produces $\Omega^{(0)}_q$ with one boundary
more. In the Quartic Kontsevich model the $g=0$ free energy
has an expression as a residue
at the poles $a$ of $\omega_{0,1}(z)$:
\begin{align*}\label{f0}
  \mathcal{F}^{(0)} = \frac{1}{2} \sum_a \biggl [ \Res\displaylimits_{q \to a}
  \omega_{0,1}(q)V_a(q) +t_a \mu_a \biggl ] -\frac{\lambda}{2N}\sum_k r_k e_k^2 +
\frac{\lambda^2}{2N^2} \sum_{k,i, k \neq i} r_ir_k \log(e_i-e_k) \\
  \quad \mu_a:= \lim_{q \to a} \big(
  V_a(q) -t_a\log[\xi_a(q)]- \int \omega_{0,1} \big)\;.
\end{align*}
 (see \cite{Branahl:2020uxs} for the definition of the potential $V_a$, the temperatures $t_a$ and the local variables $\xi_a(q)$). We give in Fig. \ref{fig:freeenergy} the elements of $\mathfrak{G}^{0,v}_\emptyset$ up to $v=2$.
Let us pick the order $\lambda^1$ to visualise the perturbative action
of the boundary creation operator. The creation
operator cuts all strands of the ribbon graph giving rise to two
different graphs (two equal by symmetry, explaining the order 2 of the
automorphism group) with one boundary and one loop:

\begin{figure}[!h]
  \vspace*{-1.8cm}
	\begin{tikzpicture}[scale=1.85, line width=\lwMacro, font=\normalsize, miter limit=\miterLimit, baseline]
	
	\def\legX{7}
	\def\legY{-3.5}
	\def\abs{-0.25}
	\def\len{0.5}
	\begin{scope}[scale=\smallScale+0.2, shift={(\legX,\legY)}]
		\draw[q] (0,0) -- (\len,0) node[right] {q};
		\draw[n] (0,\abs) -- (\len,\abs) node[right] {n};
		\draw[k] (0,2*\abs) -- (\len,2*\abs) node[right] {k};
		\draw[l] (0,3*\abs) -- (\len,3*\abs) node[right] {l};
		\draw (0.9,0.2) rectangle (-0.2,-1);
	\end{scope}



	\def\posUp{-0.55}
	\def\posDown{0}
	
	\def\i{-2}
	\node[draw] at (\posZero,\i+0.75) {$\lambda^1$};
	\ZeroOne{(\posZero,\i)}{n}{k}{l}
	\draw[cut] (0.25,\i) -- (0.25,\i+0.4) node[above] {\RNum{1}};
	\draw[cut] (0.75,\i) -- (0.75,\i+0.4) node[above] {\RNum{2}};
	\node[right=\eqGraphs, scale=\mathScale] at (\posDown,\i) {$ \displaystyle \sim \repNKL\qquad \qquad  \frac{1}{(e_n{+}e_k)(e_n{+}e_l)} $};
	
	\def\i{-3}
	\node[red,left] at (\resZero-\numGap,\i) {\RNum{1}};
	\coordinate (down) at (\resZero,\i-\BraceWidth);
	\coordinate (up) at (\resZero,\i+\BraceWidth);
	\draw[klammer] (down) node[above right] {cut n $\to$} -- (up) node[below right] {cut k $\to$};
	\scoped[scale=\smallScale] \OneOne{($(up)+(\posMiniGraph,\posUp)$)}{q}{n}{l};
	\scoped[scale=\smallScale] \OneOne{($(down)+(\posMiniGraph,\posDown)$)}{k}{q}{l};
	\coordinate (down) at (\resZero+\braceGap,\i-\BraceWidth);
	\coordinate (up) at (\resZero+\braceGap,\i+\BraceWidth);
	\draw[klammer] (up) -- (down);
	
	\def\i{-3}
	\node[red,left] at (\resOne-\numGap,\i) {\RNum{2}};
	\coordinate (down) at (\resOne,\i-\BraceWidth);
	\coordinate (up) at (\resOne,\i+\BraceWidth);
	\draw[klammer] (down) node[above right] {cut n $\to$} -- (up) node[below right] {cut l $\to$};
	\scoped[scale=\smallScale] \OneOne{($(up)+(\posMiniGraph,\posUp)$)}{q}{n}{k};
	\scoped[scale=\smallScale] \OneOne{($(down)+(\posMiniGraph,\posDown)$)}{l}{q}{k};
	\coordinate (down) at (\resOne+\braceGap,\i-\BraceWidth);
	\coordinate (up) at (\resOne+\braceGap,\i+\BraceWidth);
	\draw[klammer] (up) -- (down);
	
	\def\i{-4}
	\coordinate (down) at (\resZero+\braceGap,\i-\BraceWidth);
	\coordinate (up) at (\resZero+\braceGap,\i+\BraceWidth);
	\node[right=\eqGraphs,scale=\mathScale] at (\posZero,\i) {$ \displaystyle \sim 2 \times \repNK \frac{1}{(e_q{+}e_n)^2} \bigg(\frac{1}{e_n{+}e_k} + \frac{1}{e_q{+}e_k} \bigg) $};
 
\end{tikzpicture}
\caption{The action of the boundary creation operator for the genus zero
  free energy at $\mathcal{O}(\lambda^1)$: there are four ways to cut
  the ribbons, however only two are distinct -- explaining pictorially
  the order 2 of the automorphism group of this graph. One obtains
  the $\mathcal{O}(\lambda^1)$-contribution to the planar $2$-point
  function.}
\end{figure} 

\noindent
One can perform exactly the same technique at $\mathcal{O}(\lambda^2)$
giving the graphs in Fig.~\ref{fig:2PGraphs}.

We see that the initial data $\Omega^{(0)}_{q_1}(z)$ have to be
identified with the partially summed $2$-point function.
Repetitive application of the creation operator gives the ribbon graph
expansion of the higher $\Omega^{(g)}_{q_1,...,q_m}$.

\subsection{Enumeration of graphs}

A natural question might be the
following: Could one give a closed form for the numbers of graphs
contributing to the cumulants and $\Omega^{(g)}_{q_1,...,q_m}$ at a
certain order? In this subsection we give a partial answer
by exploiting the duality
(interchanging the r\^ole of vertices and faces)
between ribbon graphs and maps on surfaces with marked faces.

Observing that the Hermitian 1-matrix model is governed by topological
recursion caused an enormous progress in the enumeration of maps with
$k$-angulations and $m$ marked faces of any lengths (explained in a
very readable way in \cite{Eynard:2016yaa}). Only allowing for a
quartic potential, a connection to the correlation functions of the
Quartic Kontsevich Model seems natural. However, despite having the
same partition function if one sets $d=1$ (an $N$-fold degenerate
eigenvalue $e_1=e$, giving the same weight to every strand in the
ribbon graph), one has to look carefully at the definition of the
objects studied in the original works, namely cumulants like
$\langle \prod_{i=1}^b \mathrm{Tr}\, \phi^{L_i} \rangle_c$ for a
sequence $(L_i)_{i=1}^b$ of natural numbers. Luckily, a very recent
investigation \cite{Borot:2021eif} showed that when exchanging the
role of $x$ and $y$ as the ingredients of the spectral curve of the
topological recursion in the 1-matrix model one counts the so-called
\textit{fully simple maps}. These are equivalent to our correlation
functions $G^{(g)}_{|k_1^1...k_{n_1}^1|...|k_1^b...k_{n_b}^b|}$ when
replacing the traces (summed over all indices, possibly equal) by a
set of $b$ pairwise disjoint cycles $\gamma_i=(k^i_1,...,k_{n_i}^i)$
of length $l(\gamma_i)=n_i$. This formulation takes pairwise different
indices into account.

Therefore, only a subset of the \textit{ordinary} graphs/maps from the
former investigation of the 1MM can be generated. These fully simple
maps can be concretely characterised by allowing only boundaries where
no more than two edges belonging to the boundary are incident to a
vertex. The enumeration of this kind of maps was investigated for
quadrangulations by Bernardi and Fusy
\cite{bernardi2017bijections}. Building on their results, we can
relate (the $\lambda$-expansion of) our correlation functions
$G^{(0)}_{|k_1^1...k_{n_1}^1|...|k_1^b...k_{n_b}^b|}$ ($n_i$ even)
for $d=1$ to the 
number of (planar) fully simple maps/ribbon graphs: 
 \begin{align*}
   G^{(0)}_{|k_1^1...k_{n_1}^1|...|k_1^b...k_{n_b}^b|}
   \Big|_{d=1}
   =\sum_{n=0}^{\infty} \frac{3^{b+n-2}(\#n_e-1)!}{n!(3l_h+b+n-2)!}
   \prod_{i=1}^b n_i \binom{\frac{3n_i}{2}}{\frac{n_i}{2}}
   \cdot \frac{(-\lambda)^{n+l_h+b-2}}{(2e)^{2(n+l_h+b-1)}}\;.
\end{align*}
Here $l_h:=\frac{1}{2}\sum_i n_i$ is the half boundary length and
$\#n_e:=3l_h+2b+2n-4$ the number of edges.  For  $b=1$, $n_1=2$ and thus $l_h= 1$, one recovers  the famous
result of Tutte for the number of rooted planar quadrangulations
$\frac{2 \cdot 3^n(2n)!}{n!(n+2)!}$, obtained as coefficient of
$\frac{(-\lambda)^n}{(2e)^{2n+2}}$ in the 2-point
function. Compare the first terms of this sequence $(1,2,9,54,...)$ with
Fig.~\ref{fig:2PGraphs}. Once again, we can apply the creation
operator to the $d=1$ result of $\mathcal{F}^{(0)}$ being
 \begin{align*}
& \mathcal{F}^{(0)}=  \sum_{n=1}^{\infty}  3^n  \frac{(2n-1)!}{n!(n+2)!}   \frac{(-\lambda)^n}{(2e)^{2n}} =\frac{-\lambda}{2(2e)^2}+\frac{9(-\lambda)^2}{8(2e)^4} +\frac{9(-\lambda)^3}{2(2e)^6} +\frac{189(-\lambda)^4}{8(2e)^8}  +...
\end{align*}
For example, the orders $2,2,8$ of the automorphism groups the graphs
contributing to 
$\mathcal{O}(\lambda^2)$ in Fig.~\ref{fig:freeenergy} produce the coefficient 
$\frac{1}{2}+\frac{1}{2}+\frac{1}{8}=\frac{9}{8}$
in front of $\frac{(-\lambda)^2}{(2e)^4}$. 
Acting with $-\frac{\partial}{\partial e}$ on $\mathcal{F}^{(0)}$
gives the aforementioned numbers
$\frac{2 \cdot 3^n(2n)!}{n!(n+2)!}$, providing again evidence for the
correctness of the creation operator.

We give a very brief outlook on the transition from
$ G^{(g)}_{|k_1^1...k_{n_1}^1|...|k_1^b...k_{n_b}^b|}$ (fully simple
quadrangulations) to $\Omega_{q_1,...,q_n}^{(g)}$. Currently, we know
that $\Omega_{q}^{(g)}$ generate the number of \textit{ordinary}
quadrangulations as in the Hermitian one-matrix model. The basic
relation
$\Omega^{(g)}_{q} := \frac{1}{N}\sum_{k=1}^N
G^{(g)}_{|qk|}+\frac{1}{N^2}G^{(g)}_{|q|q|}$ extends the relation for
$g=1$ found in \cite{Borot:2017agy} to any genus $g$. Moreover, we
remark that the pure TR constituents of $\Omega_{q}^{(g)}$ seem to be
a generating function of only the \textit{bipartite} rooted
quadrangulations (so far known for $g=0,1,2$).

An interpretation of all $\Omega_{q_1,...,q_n}^{(g)}$, $n>1$, with and without blobs, especially in a closed or recursive form, is under current investigation.

\section{Summary and outlook}

In this article we reviewed the accomplishments of the last two
decades towards the exact solution of a scalar quantum field theory 
on noncommutative geometries of various dimensions. We highlighted how
the recursive structure in the solution theory fits into a larger
picture in complex algebraic geometry.
After having introduced our understanding of quantum
fields on a noncommutative space as well as the powerful machinery of
(blobbed) topological recursion, we mainly focused on results of the
previous three years: With the discovery of meromorphic differentials
in our model which are governed by blobbed topological recursion, the
long-term project of its exact solution seems to gradually come to an
end. We are convinced of having found the most suitable structures
that lead us to the path of a complete understanding of the underlying
recursive patterns in the quantum field theory.

We interpret this plethora of fascinating algebraic structures as the
main reason for the exact solvability of this class of
QFT-models. This is highly exceptional. We started with the most
concrete results in the finite-matrix (zero-dimensional) case, the
quartic Kontsevich model, and explained how it relates to structures
in algebraic geometry such as ramified coverings of Riemann surfaces,
meromorphic differential forms and the moduli space of complex curves.
Then we explained how at in the simplest topological sector one can
achieve the limit to a quantum field theory models on 2- and
4-dimensional Moyal space. The most remarkable result here is the
absence of any triviality problem.  We concluded with a glimpse into
perturbation theory, which we not only see as a useful cross-check of
our exact results, but also as an active area of research with
interesting connections to enumerative geometry and number theory.

The mind-blowing algebro-geometric tool of topological recursion shall
lead us one day to the long awaited property of integrability of a
quantum field theory. To reach this goal at the horizon and also to
explore further structures and phenomena, many questions are still on
the agenda. We finally list an excerpt:
\begin{itemize}
\item Does a generic structure of the holomorphic part of
  $\omega_{g,n}$ also exist in the non-planar sector? How does it look
  like (also taking the free energies into consideration)?

\item The recursion formula of the holomorphic add-ons shares many
  characteristics with usual topological recursion. Can we achieve
  these results also with pure TR by changing the spectral curve
  (e.g.\ of genus 1)?
  
\item How can we formulate BTR in 4 dimensions, where no algebraic
  ramification point is given anymore?

\item Does the particular combination of Feynman graphs contributing to 
$\omega_{g,n}$ have a particular meaning in quantum field theory?

\item Which property of the moduli space of stable complex curves is
  captured by the intersection numbers generated by the quartic
  Kontsevich model?
  
\item Is there an integrable hierarchy behind our quantum field theory?
\end{itemize}
We are looking forward to many interesting insights in the
not-too-distant future and invite the reader to follow our
progress towards blobbed topological recursion of a noncommutative
quantum field theory!

\section*{Acknowledgements}

Our work was supported\footnote{``Funded by
  the Deutsche Forschungsgemeinschaft (DFG, German Research
  Foundation) -- Project-ID 427320536 -- SFB 1442, as well as under
  Germany's Excellence Strategy EXC 2044 390685587, Mathematics
  M\"unster: Dynamics -- Geometry -- Structure.''} by the Cluster of
Excellence \emph{Mathematics M\"unster} and the CRC 1442 \emph{Geometry:
  Deformations and Rigidity}. AH is supported through
the Walter-Benjamin fellowship\footnote{``Funded by
  the Deutsche Forschungsgemeinschaft (DFG, German Research
  Foundation) -- Project-ID 465029630}.


\end{document}